\documentclass[apjl]{emulateapj}
\usepackage{amsfonts,amsmath,graphicx,natbib,apjfonts,subfigure,color,gensymb,longtable}

\def\gw{GW170817A}
\def\nuu{1}
\def\cfa{2}
\def\nyu{3}
\def\columbialorenzo{4}
\def\columbia{5}
\def\purdue{6}
\def\hf{7}
\def\fr{8}
\def\oh{9}

\def\chi{18}
\shorttitle{\gw\, a hundred days after merger}
\shortauthors{Margutti~et~al.}

\begin{document}
\title{The Binary Neutron Star  event LIGO/VIRGO GW170817 a hundred and sixty days after merger: synchrotron emission across the electromagnetic spectrum}

\author{R.~Margutti\altaffilmark{\nuu}, 
K.~D.~Alexander\altaffilmark{\cfa},
X. Xie\altaffilmark{\nyu},
L.~Sironi\altaffilmark{\columbialorenzo},
B.~D.~Metzger\altaffilmark{\columbia},
A.~Kathirgamaraju\altaffilmark{\purdue},
W.~Fong\altaffilmark{\nuu}$^,$\altaffilmark{\hf},
P.~K.~Blanchard\altaffilmark{\cfa},
E.~Berger\altaffilmark{\cfa}, 
A. MacFadyen\altaffilmark{\nyu},
D. Giannios\altaffilmark{\purdue},
C.~Guidorzi\altaffilmark{\fr},
A.~Hajela\altaffilmark{\nuu},
R.~Chornock\altaffilmark{\oh},
P.~S.~Cowperthwaite\altaffilmark{\cfa},
T.~Eftekhari\altaffilmark{\cfa},
M.~Nicholl\altaffilmark{\cfa},
V.~A.~Villar\altaffilmark{\cfa},
P.~K.~G.~Williams\altaffilmark{\cfa},
J. ~Zrake\altaffilmark{\columbia}
}

\altaffiltext{\nuu}{Center for Interdisciplinary Exploration and Research in Astrophysics (CIERA) and Department of Physics and Astronomy, Northwestern University, Evanston, IL 60208}
\altaffiltext{\cfa}{Harvard-Smithsonian Center for Astrophysics, 60 Garden Street, Cambridge, MA 02138, USA}
\altaffiltext{\nyu}{Center for Cosmology and Particle Physics, New York University, 726 Broadway, New York, NY 10003, USA}
\altaffiltext{\columbialorenzo}{Columbia University, Pupin Hall, 550 West 120th Street, New York, NY 10027, USA}
\altaffiltext{\columbia}{Department of Physics and Columbia Astrophysics Laboratory, Columbia University, New York, NY 10027, USA}
\altaffiltext{\purdue}{Department of Physics and Astronomy, Purdue University, 525 Northwestern Avenue, West Lafayette, IN 47907, USA}
\altaffiltext{\hf}{Hubble Fellow}
\altaffiltext{\fr}{Department of Physics and Earth Science, University of Ferrara, via Saragat 1, I--44122, Ferrara, Italy}
\altaffiltext{\oh}{Astrophysical Institute, Department of Physics and Astronomy, 251B Clippinger Lab, Ohio University, Athens, OH 45701, USA}

\begin{abstract}
We report deep Chandra, HST and VLA observations of the binary neutron star event GW170817 at $t<160$ d after merger. These observations show that GW170817 has been steadily brightening with time and might have now reached its peak, and constrain the emission process as non-thermal synchrotron emission where the cooling frequency $\nu_c$ is above the X-ray band and the synchrotron frequency $\nu_m$ is below the radio band.  The very simple power-law spectrum extending for eight orders of magnitude in frequency enables the most precise measurement of the index $p$ of the distribution of non-thermal relativistic electrons $N(\gamma)\propto \gamma^{-p}$ accelerated by a shock launched by a NS-NS merger to date. We find $p=2.17\pm0.01$, which indicates that radiation from ejecta with $\Gamma\sim3-10$ dominates the observed emission. While constraining the nature of the emission process, these observations do \emph{not} constrain the nature of the relativistic ejecta. We employ simulations of explosive outflows launched in NS ejecta clouds to show that the spectral and temporal evolution of the non-thermal emission from GW170817  is consistent with both emission from radially stratified quasi-spherical ejecta traveling at mildly relativistic speeds, \emph{and} emission from off-axis collimated ejecta characterized by a narrow cone of ultra-relativistic material with slower wings extending to larger angles. In the latter scenario, GW170817 harbored a normal SGRB directed away from our line of sight.  Observations 
at $t\le 200$ days are unlikely to settle the debate as in both scenarios the observed emission is effectively dominated by radiation from mildly relativistic material. 
\end{abstract}
\section{Introduction}
\label{Sec:intro}
The joint discovery of gravitational waves \citep{Abbott17discovery} and photons
from the first binary neutron star (BNS) merger event GW170817  established that gravitational-wave detected BNS mergers can be accompanied by detectable emission across the electromagnetic spectrum, 
including $\gamma$-rays \citep{Goldstein17,Savchenko17}. During the first $\sim 15$ days the spectrum consisted of a combination of thermal emission powered by the radioactive decay of  heavy elements freshly synthesized in the merger ejecta (i.e. the kilonova emission, KN; \citealt{Metzgerhandbook,Chornock17,Coulter17,Cowperthwaite17,Drout17,Kasliwal17,Nicholl17,Pian17,Smartt17,Soares-Santos17,Tanvir17,Valenti17,Villar17}) and non-thermal synchrotron emission dominating in the X-rays and radio bands \citep{Alexander17GW, Haggard17GW,Hallinan17,Margutti17GW,Troja17GW}. The thermal component later subsided. During $\sim160$ days of intense monitoring, the non-thermal emission brightened with time \citep{Mooley17,Ruan17,MarguttiAtellast,TrojaGCNlast,Troja18} and might have now reached its peak  as we show below (see also \citealt{Davanzo18, Troja18GCN}). The most pressing question regards the intrinsic nature of GW170817. 

A first possibility is that GW170817 is an intrinsically sub-luminous event with total gamma-ray energy released $E_{\gamma,iso}\sim6\times 10^{46}\,\rm{erg}$. As a comparison, classical cosmological Short Gamma-Ray Bursts (SGRBs) typically have $E_{\gamma,iso}\sim 10^{50}-10^{52}\,\rm{erg}$  \citep{Fong15,Berger14}. In this scenario, GW170817 did not produce a successful collimated relativistic outflow   (i.e. no observer in the Universe observed a classical SGRB in association with GW170817), the emission from GW170817 is quasi-spherical and powered by energy deposited by the interaction of the unsuccessful jet with the BNS ejecta (\citealt{Gottlieb17}). The simplest incarnation of this model (i.e. the uniform fireball)  fails to reproduce current observations, but a more complex version with highly stratified ejecta with energy  $E(>\Gamma \beta)\propto (\Gamma\beta)^{-5}$ (where $\Gamma\beta$ in this context is the specific momentum of the outflow) successfully accounts for the observed properties of GW170817. This model was favored by \cite{Gottlieb18,Hallinan17,Kasliwal17,Mooley17,Nakar18}.

\tabletypesize{\small}
\begin{deluxetable*}{lcccc}
\tablecolumns{5}
\tablewidth{0pc}
\tablecaption{X-ray Spectral Parameters and inferred flux ranges ($1\,\sigma$ c.l.). Upper limits are provided at the $3\,\sigma$ c.l. 
\label{tab:spec}}
\tablehead {
\colhead {Obs ID} &  \colhead {Time since merger } &  \colhead {$\Gamma$} &   \colhead {Flux (0.3-10 keV)}     &   \colhead {Unabsorbed Flux (0.3-10 keV)}         \\  
                    &   \colhead {(days)}       &                                      &   \colhead {($10^{-15}\rm{erg\,s^{-1}cm^{-2}}$)}  &  \colhead {($10^{-15}\rm{erg\,s^{-1}cm^{-2}}$)} 
}
\startdata
18955 & 2.34  &1.4  & $<1.8^{a}$ & $<1.9^{a}$\\
19294 & 9.21 &$0.95^{+0.95}_{-0.19}$ & $(4.2- 9.3)^{b}$  & $(4.4- 9.6)^{b}$\\
	   &  & $1.4^{+0.9}_{-0.1}$ & $(2.7 - 6.8)^{d}$ & $(2.9-7.3)^{d}$ \\
20728 & 15.39 &$1.6^{+1.5}_{-0.1}$ & $(3.0- 5.6)^{c}$ & $(3.1- 5.8)^{c}$\\
	  &   & $1.4^{+0.9}_{-0.1}$ &  $(3.7- 7.3)^{d}$ & $(4.0-7.8)^{d}$\\
18988    &15.94& $1.4$  &  $(3.8- 7.5)^{e}$  & $(4.1- 8.0)^{e}$\\
20860/1 & 109.39&$1.62^{+0.16}_{-0.16}$ & $(20.-25.)^{b}$ & $(22.-28.)^{b}$\\
20936/7/8/9-20945 & 158.50& $1.61^{+0.17}_{-0.17}$ & $(22.-27.)^{b}$&$(24.-29.)^{b}$ \\ 
\enddata
\tablecomments{$^{a}$ 0.5-8 keV count-rate upper limit  of  $1.2\times 10^{-4}\,\rm{cps}$ from \cite{Margutti17GW}, with updated flux calibration performed with an absorbed power-law model with $\Gamma=1.4$ as inferred from our joint fit of the CXO observations with IDs 19294 and 20728. 
$^{b}$ This work. \\
$^{c}$  From \cite{Margutti17GW}. \\
$^{d}$  From a joint spectral fit of CXO observations, IDs 19294 and 20728. This work.\\
$^{e}$  Flux from \cite{Haggard17GW} re-scaled to the $\Gamma=1.4$ spectrum. This work.}
\end{deluxetable*}

Here we present deep radio, optical and X-ray observations of GW170817 $\sim110-160$ d after merger (Sec. \ref{Sec:Obs}) and offer an alternative interpretation. We employ hydrodynamical simulations of the jet interaction with the BNS ejecta to show that a core of ultra-relativistic material can successfully break through the closest environment and power a classical SGRB in association with GW170817,  in agreement with the recent results by \cite{Lazzati17post,LazzatiNature17}. We further demonstrate in Sec. \ref{Sec:int} that the very simple power-law spectrum extending for  eight orders of magnitude in frequency allows a precise measure of the properties of electrons accelerated at the shock front. In particular it enables inferences on the slope of the non-thermal tail of accelerated particles from which we derive robust constraints on the shock velocity which are independent from the morphology of the outflow (collimated vs. spherical). We demonstrate that all these properties are consistent with a SGRB-like outflow originally directed away from our line of sight (Sec. \ref{Sec:int}). In this scenario GW170817 is not intrinsically subluminous and its unusual observed properties result from a different viewing angle than classical SGRBs, which are viewed along the jet axis.
We conclude in Sec. \ref{Sec:Conc}.

We assume that all electrons are shock accelerated to a power-law energy distribution $N(\gamma)\propto \gamma^{-p}$, i.e. $\xi_N=1$, which is the standard assumption in GRB studies. If only a fraction  of electrons $\xi_N<1$ is accelerated into the non-thermal tail, the inferred density should be re-scaled as $n/\xi_N$. We adopt the convention $F_{\nu}\propto \nu^{-\beta}$ and $\Gamma=\beta+1$, where $\beta$ is the spectral index and $\Gamma$ is the photon index. We assume a distance to NGC\,4993 of 39.5\,Mpc ($z=0.00973$) as listed in the NASA Extragalactic Database. $1\,\sigma$ c.l. uncertainties are listed unless otherwise stated. In this manuscript we employ the notation $Q_x\equiv Q/10^x$.

\section{Observations and data analysis}
\label{Sec:Obs}
\subsection{Chandra X-ray Observations}
\label{SubSec:Chandra}

\tabletypesize{\small}
\begin{deluxetable*}{ccccl}
\tablecolumns{5}
\tablewidth{0pc}
\tablecaption{VLA observations of GW170817. 
\label{tab:radio}}
\tablehead {
\colhead {Time since merger } &  Mean Freq & Freq Range & On-source &\colhead {Flux Density}       \\  
  \colhead {(days)}       &            (GHz)    &    (GHz)    & Time (hr)             &   \colhead {($\mu Jy$)} 
}
\startdata
 80.10 & 6.0 & $3.976-7.896$ & $1.5$ & $37.4\pm4.2$ \\ 
 112.04 & 5.0 & $3.796-5.896$ & $1.5$ & $69.7 \pm 7.5$ \\ 
 112.04 & 7.0 & $5.976-7.896$ & $1.5$ & $57.7 \pm 4.7$ \\ 
 115.05 & 2.6 & $2.088-2.984$ & $0.57$ & $82.3 \pm 20.7$ \\ 
 115.05 & 3.4& $2.888-3.784$ & $0.57$ & $95.8 \pm 11.0$ \\ 
 115.05 & 9.0 & $7.976-9.896$ & $0.69$ & $56.4 \pm 10.4$ \\
 115.05 & 11.0 & $9.976-11.896$ & $0.69$ & $52.5\pm 10.1$ \\
 115.05 & 13.0 & $11.976-13.896$ & $1.59$ & $42.3 \pm 5.7$ \\
 115.05 & 15.0 & $13.976-15.896$ & $1.59$ & $45.2\pm 7.0$ \\
 115.05 & 17.0 & $15.976-17.896$ & $1.59$ & $44.0\pm 7.9$ \\
 162.89 &  2.6  &$2.088 - 3.016$& $0.58$ & $104.5\pm22.3$\\
 162.89 &  3.4  &$3.016 - 3.912$& $0.58$ & $91.2	\pm17.4$\\
 162.89 &  5.0  &$4.000 - 6.000$& $0.70$ & $80.8\pm	12.5$\\
 162.89 &  7.0  &$6.000 - 8.000$& $0.70$ & $61.1\pm	7.3$\\
 162.89 &  9.0  &$8.000 - 10.000$& $0.70$ & $55\pm	9.9$ \\
 162.89 &  11.0  &$10.000 - 12.000$& $0.70$ & $34.4\pm	10.$\\
 162.89 & 13.0   &$12.000 - 14.000$& $1.84$ & $41.7	\pm6.3$\\
 162.89 &  15.0  &$14.000 - 16.000$&$1.84$ & $38.9\pm	7.2 $\\
 162.89 &  17.0  &$16.000 - 18.000$&$1.84$ & $43.5\pm	7.7$
\enddata
\end{deluxetable*}

We observed GW170817 with the \emph{Chandra} X-ray Observatory (CXO) on 2017 August 19.71UT,  $\delta t\approx 2.3\,\rm{d}$ after the GW trigger (observation ID 18955; PI: Fong; Program 18400052), leading to a deep X-ray non-detection with $L_x<3.2\times 10^{38}\,\rm{erg\,s^{-1}}$ (\citealt{Margutti17GW}) that sets GW170817 apart from all previous SGRBs seen on-axis (\citealt{Fong17GW}). Further CXO observations obtained at $\delta t\approx 9\,\rm{d}$ (\citealt{Troja17GW}, observation ID 19294; PI: Troja; Program 18500489) and $\delta t\approx 15\,\rm{d}$ (\citealt{Haggard17GW, Margutti17GW,Troja17GW}, observation IDs 18988, 20728; PIs: Haggard, Troja; Programs 18400410,18508587) since merger revealed X-ray emission at the location of GW170817 with rising temporal behavior. 

We independently re-analyzed the CXO observations acquired $\delta t\approx 9\,\rm{d}$ post-merger (ID 19294) and originally presented in \cite{Troja17GW}.
 \emph{Chandra} ACIS-S data have been reduced with the {\tt CIAO} software package
(v4.9) and relative calibration files, applying standard ACIS data filtering as in \citealt{Margutti17GW}.  Using {\tt wavdetect} we find that an X-ray source is clearly detected with significance of 5.8$\,\sigma$ at the location of the optical counterpart of GW170817.  The inferred count-rate in the 0.5-8 keV energy range is  $(2.9\pm0.8)\times 10^{-4}\,\rm{c\,s^{-1}}$ (exposure time of 49.4 ks), consistent with the results from  \cite{Troja17GW}. 
We employ Cash statistics to fit the spectrum. We adopt an absorbed power-law spectral model with index $\Gamma$ and Galactic neutral hydrogen column density $\rm{NH}_{mw} = 0.0784\times10^{22}\,\rm{cm^{-2}}$ (\citealt{Kalberla05}) and use MCMC sampling to constrain the spectral parameters. We find $\Gamma=0.95^{+0.95}_{-0.19}$. We find no statistical evidence for intrinsic neutral hydrogen absorption and place a limit $\rm{NH_{int}}<7 \times 10^{22}\,\rm{cm^{-2}}$ ($3\,\sigma$ c.l.). For these parameters the 0.3-10 keV  flux is $(4.2- 9.3)\times10^{-15}\,\rm{erg\,s^{-1}cm^{-2}}$ ($1\,\sigma$ c.l.), corresponding to an unabsorbed flux of $(4.4- 9.6)\times10^{-15}\,\rm{erg\,s^{-1}cm^{-2}}$.

Comparison with the X-ray spectrum of GW170817 at $\delta t\approx 15\,\rm{d}$ (ID 20728) that we presented in \citealt{Margutti17GW} indicates a possibly harder emission at early times ($\Gamma=0.95^{+0.95}_{-0.19}$ at $\delta t\approx 9\,\rm{d}$ vs. $\Gamma=1.6^{+1.5}_{-0.1}$  at  $\delta t\approx 15\,\rm{d}$). While we find this possibility intriguing, the limited number statistics of the two spectra does not allow us to draw conclusions as the two $\Gamma$ values are statistically consistent.  A joint spectral fit of the two epochs indicates $\Gamma=1.4^{+0.9}_{-0.1}$ ($1\,\sigma$ c.l.) with a $3\sigma$ upper limit $\rm{NH_{int}}<2.7 \times 10^{22}\,\rm{cm^{-2}}$. The corresponding flux ranges are reported in Table \ref{tab:spec}. Our results from the joint fit are broadly consistent with the findings from \cite{Troja17GW}. 

Deep X-ray observations of GW170817 have been obtained as soon as the source re-emerged from Sun constraint (PI Wilkes, observation IDs 20860, 20861; Program 18408601; \citealt{MarguttiAtellast, MarguttiGCNlast,HaggardGCNlast,TrojaGCNlast}). The CXO started observing GW170817 on 2017 December 3.07UT  ($107.5$ d since merger, ID 20860) for 74.1 ks. An X-ray source is clearly detected at the location of GW170817 with significance of $33.4\,\rm{\sigma}$ and net count-rate $(1.47\pm0.14)\times 10^{-3}\,\rm{c\,s^{-1}}$ (0.5-8 keV). The CXO observed the field for an additional 24.7 ks starting on 2017 December 6.45UT ($110.9$ d since merger,  ID 20861). The X-ray source is still detected with a significance of $\sim 15.0\,\sigma$ and net count-rate of $(1.41 \pm 0.24)\times 10^{-3}$ (0.5-8 keV).  The joint spectrum can be fit with an absorbed power-law spectral model with photon index $\Gamma= 1.62 \pm 0.16$ (1 sigma c.l.), consistent with the results from \cite{Ruan17}.  We find no evidence for intrinsic neutral hydrogen absorption and constrain $\rm{NH_{int}}<0.7\times 10^{22}\,\rm{cm^{-2}}$ ($3\,\sigma$ c.l.).  These properties are consistent with the X-ray spectral properties of GW170817 at $t\le15$ days. The 0.3-10 keV inferred flux range is  $(2.0-2.5)\times 10^{-14}\,\rm{erg\,s^{-1}cm^{-2}}$, (unabsorbed flux of $(2.2-2.8)\times 10^{-14}\,\rm{erg\,s^{-1}cm^{-2}}$). This result indicates substantial brightening of the X-ray source during the last $\sim95$ d with no measurable spectral evolution (Fig. \ref{Fig:SED}). 

Further CXO observations have been obtained between 2018 January 17 and 28, 153.4-163.8 d since merger (PI Wilkes, observation IDs 20936, 20937, 20938, 20939, 20945; Program 19408607, total exposure time of 104.8 ks). GW170817 is detected with high confidence in each observation. The total source count-rate is $157.1 \pm 12.7 $ (0.5-8 keV), corresponding to $(1.50\pm0.12)\times 10^{-3}\,\rm{c\,s^{-1}}$. We do not find evidence for statistically significant spectral evolution during the entire observation.  We also do not find evidence for statistically significant temporal variability of the source during the observation. The joint spectrum can be fit with an absorbed power-law spectral model with photon index $\Gamma=1.61\pm 0.17$ and $\rm{NH_{int}}<1.0\times 10^{22}\,\rm{cm^{-2}}$ ($3\,\sigma$ c.l.). These results are broadly consistent with the preliminary analysis by  \cite{Troja18GCN} and \cite{Haggard18}. The corresponding 0.3-10 keV observed flux range is  $(2.2-2.7)\times 10^{-14}\,\rm{erg\,s^{-1}cm^{-2}}$, while the unabsorbed flux is  $(2.4-2.9)\times 10^{-14}\,\rm{erg\,s^{-1}cm^{-2}}$. This result indicates that the source did not experience significant temporal and spectral evolution between $\sim100$ d and $\sim150$ d since merger. Our findings do not support the claim of declining emission from GW170817 by \cite{Davanzo18}, but suggest that the non-thermal emission from GW170817 is now close to its peak.

\subsection{HST Observations}
\label{SubSec:HST}

We obtained 1 orbit of \textit{Hubble Space Telescope} (\textit{HST}) observations of GW170817 on 1 January 2018 (137 d since merger) using the Advanced Camera for Surveys (ACS) with the F606W filter (PID: 15329; PI: Berger).  We produced a drizzled image corrected for optical distortion using the {\tt astrodrizzle} task in the {\tt drizzlepac} software package provided by STScI.  We detect a faint source at the location of the optical counterpart of GW170817, confirmed by relative astrometry with our ACS/F625W image from 27 August 2017 \citep{Cowperthwaite17}.  To measure the flux of the source we first subtract a model of the galaxy surface brightness profile determined using GALFIT v3.0.5 \citep{Peng10}.  Using aperture photometry and the ACS/F606W zeropoint provided by the \textit{HST} team, we find an observed AB magnitude of $26.90 \pm 0.25$ mag.  Correcting for Galactic 
extinction with $E(B-V) = 0.105$ mag \citep{Schlafly11}, the extinction corrected AB magnitude is $26.60 \pm 0.25$ mag.
As a comparison, at 110 d since merger,  \cite{Lyman18} find $m=26.44\pm0.14$ mag.
\subsection{VLA Observations}
\label{SubSec:VLA}

Our radio observations of GW170817 from $0.5-39$ d since merger have been reported in \cite{Alexander17GW}. We continued observing GW170817 with the Karl J. Jansky Very Large Array (VLA) under program 17A-231 (PI: Alexander), obtaining observations on 5 November 2017 ($\delta t\sim 80$ d since merger) at a mean frequency of 6 GHz (C band), using a bandwidth of 4 GHz. These new observations were taken in the VLA's B configuration. We analyzed and imaged the VLA data using standard CASA routines \citep{McMullin07}, using 3C286 as the flux calibrator and J1258$-$2219 as the phase calibrator. We fit the flux density and position of the emission using the \textsf{imtool} program within the \textsf{pwkit} package \citep{pwkit.software}. We clearly detect the source with a flux density of $37\pm4$ $\mu$Jy. The in-band spectral index is poorly constrained, but is clearly optically thin (Table \ref{tab:radio}). 

We obtained further multi-frequency VLA observations under the same program on 7 December 2017 (C band) and under program 17B-425 (PI: Alexander) on 10 December 2010 (S, X, and Ku bands, spanning the frequency range 2--18 GHz). New observations spanning 2-18 GHz (S, C, X, and Ku bands) were obtained under program 17B-425 on 27 January 2018. We reduced the data using the same procedure outlined above and cross-checked our results against the automated CASA-based VLA pipeline. The flux densities obtained with each method are fully consistent to within the error bars at all frequencies; we choose to report the pipeline flux densities here because the images have slightly lower rms noise. GW170817 is clearly detected at all radio frequencies and has continued to brighten, enabling us to split the data into narrower frequency bandwidths for imaging. At S band, we divided the data into two 1 GHz subbands, although the effective bandwidth of each after flagging is closer to 750 MHz due to RFI. At higher frequencies, we split the data into 2-GHz bandwidth. We report the measured flux densities in Table \ref{tab:radio}. As before, uncertainties were calculated using the \textsf{imtool} package and represent the uncertainty on a point source fit.  The December measurements clearly indicate an optically thin spectrum with spectral index $\beta_R=0.47 \pm 0.08$. This value is consistent with the X-ray spectral index $\beta_X=0.62 \pm 0.16$ ($\Gamma=\beta+1$) obtained a few days before (Sec. \ref{SubSec:Chandra}). The latest VLA observations in January are also optically thin with $\beta_R=0.55 \pm 0.10$, in good agreement with the CXO spectral index $\beta_X=0.621\pm 0.17$ around the same time.

\subsection{Joint X-Ray and Radio analysis}
\label{SubSec:join}
A joint spectral fit of radio data obtained at $\delta t\sim 111-114$ d and X-ray data obtained around $\delta t\approx 109$ d with a simple power-law model $F_\nu\propto \nu^{-\beta_{XR}}$ constrains $\beta_{XR}=0.588\pm0.005$. This value is consistent with the spectral indexes $\beta_{X}$ and $\beta_{R}$ derived from individual fits within the X-ray and radio bands (Sec. \ref{SubSec:Chandra},\ref{SubSec:VLA}), and shows that at $t\approx 110$ d the broad-band X-ray to radio emission from GW170817 originates from the same non-thermal spectral component. 

\begin{figure}
\center
\includegraphics[scale=0.47]{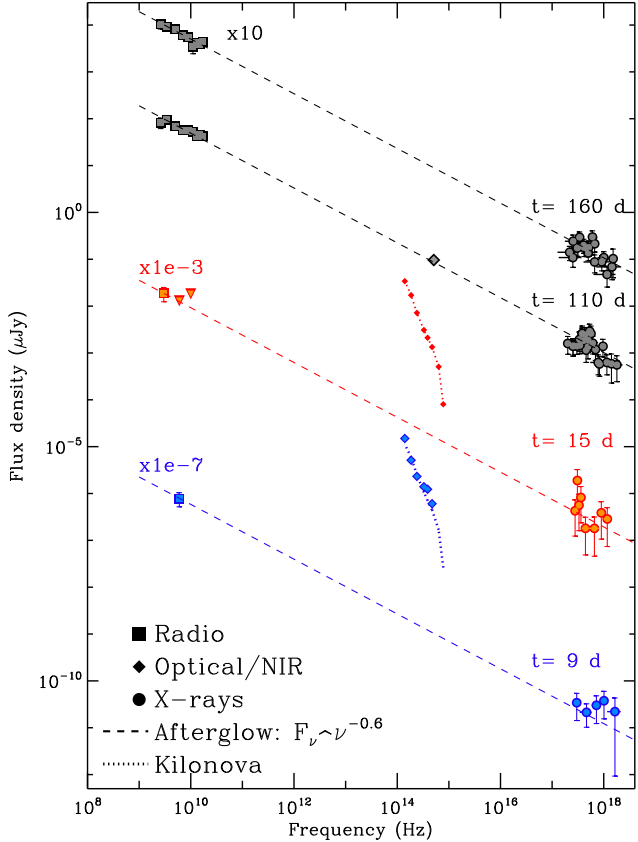}\\
\caption{Evolution of the broad-band radio-to-X-ray SED of GW170817 from 9 d until 160 d since merger. The radio and X-ray data are dominated by non-thermal synchrotron emission from the GW170817 afterglow at all times and consistently track each other on a $F_{\nu}\propto \nu^{-0.6}$ spectral power-law segment. At early times $t\le 15$ d the optical-NIR is dominated by radioactively powered emission from the KN. By day 110 the KN component has faded away and the detected optical-NIR emission is dominated by the $F_{\nu}\propto \nu^{-0.6}$ afterglow radiation.  Filled circles: CXO data. Filled squares: VLA. Note that while \cite{Hallinan17} consider their 6 GHz measurement at $\sim10$ days only as a potential detection, here we show that it does naturally lie on the $\propto \nu^{-0.6}$ extrapolation of the X-ray data, which suggests that this is in fact a real detection (and the earliest radio detection of GW170817). Filled diamonds at 15 and 9 d: optical-NIR data from \cite{Villar17}.  For day 9 we show the actual data from \cite{Tanvir17,Soares-Santos17,Cowperthwaite17,Kasliwal17}, while for day 15 we show the extrapolated values from the best fitting model from \cite{Villar17}. Black dashed line: $F_{\nu}\propto \nu^{-\beta_{XR}}$  afterglow component with $\beta_{XR}=0.6$  that best fits the observations at 110 d and 160 d. Dashed red and blue lines: same afterglow model renormalized to match the observed flux level at 15 d and 9d. Dotted line: best fitting KN component. The SED at 15 d and 9 d have been rescaled for displaying purposes. The HST observations from \cite{Lyman18} obtained at 110 d (filled diamonds) are shown here for comparison but have not been used in our fits. }
\label{Fig:SED}
\end{figure}

To refine our measurement of the X-ray to radio spectral slope $\beta_{XR}$ at $\sim110$ d we account for the (mild) temporal evolution of the afterglow flux adopting the iterative procedure that follows. We initially assume a fiducial spectral index value $\beta_i=0.60$, which is used to construct a ``master'' radio light-curve of GW170817 at a given frequency using the entire set of radio observations available at all frequencies.  Radio data have been compiled from \citet{Alexander17GW}, \citet{Hallinan17}, \cite{Kim17} and \cite{Mooley17}. We fit the master radio light-curve with a power-law model $F_{\nu}\propto t^{\alpha}$. The best-fitting $\alpha$ is then used to renormalize the flux densities measured at $\delta t=111-114$ d to a common epoch of 109 d since merger (to match the time of CXO observations). Finally, we estimate $\beta_f$ from a joint fit of the broad-band radio-to-X-ray spectrum at 109 d. This procedure is repeated until convergence (i.e. $\beta_f=\beta_f$ within error bars).  We find $\beta_{XR}= 0.585 \pm 0.005$ and $\alpha=0.73 \pm 0.04$ (Fig. \ref{Fig:SED}). As a comparison, from the analysis of radio data alone at $t<93$ d \citealt{Mooley17} infer $\beta_R=0.61\pm0.05$, consistent with our results. Our measurement of the spectral slope benefits from the significantly larger baseline of eight orders of magnitude in frequency, and is consequently more precise. We plot in Fig. \ref{Fig:SED} the HST measurement obtained by \cite{Lyman18} at 110 d.  This comparison shows a remarkable agreement with our bestfitting SED and demonstrates that at 110 d since merger the optical emission from GW170817 is of non-thermal origin and originates from the afterglow.

The X-ray and radio light-curves suggest that GW170817 might be now approaching its peak of non-thermal emission.  From a fit of the radio-to-X-ray SED at $\sim160$ d  we find $\beta_{XR}= 0.584 \pm 0.006$, consistent with the value at $110$ d. 

We compile in Fig.\ \ref{Fig:SED} the radio-to-X-ray SEDs of GW170817 at 15 d and 9 d (orange and blue symbols). At these epochs the thermal emission from the radioactive decay of freshly synthesized heavy elements  (i.e. the kilonova) dominates the UV-optical-NIR bands. Fig. \ref{Fig:SED} shows that a re-scaled version of the $\beta_{XR}= 0.585$ spectrum that best-fits the 110 d epoch adequately reproduces the X-ray and radio emission from GW170817 at all times. Interestingly, the extrapolation of the X-ray flux density at 9 d with a $\propto \nu^{-0.6}$ spectrum matches the 6 GHz measurement reported by \cite{Hallinan17} as a potential --- but possibly spurious --- detection, suggesting that the 6 GHz measurement is in fact a real detection (and the earliest radio detection of GW170817).

Based on these results we conclude that the non-thermal emission from GW experienced negligible spectral evolution across the electromagnetic spectrum in the last $\sim150$ d, and that the radio and X-ray radiation from GW170817 continue to represent the same non-thermal emission component. This component of emission is now approaching its peak.

\section{Interpretation and discussion}
\label{Sec:int}
\subsection{A synchrotron spectrum from particles accelerated by shocks with $\Gamma\approx 3-10$}
\label{SubSec:synch}
The simple power-law spectrum extending over  eight orders of magnitude in frequency indicates that radio and X-ray radiation are part of the same non-thermal emission component, which we identify as synchrotron emission. At all times of our monitoring the synchrotron cooling frequency $\nu_c$ is above the X-ray band,  $\nu_m$ is below the radio band and the observed radio and X-ray emission is on the $F_{\nu}\propto \nu^{-(p-1)/2}$ spectral segment, where $p$ is the index of the non-thermal electrons accelerated into a power-law distribution $N_e(\gamma)\propto \gamma^{-p}$ at the shock front. From our best-fitting $\beta_{XR}$, we infer $p=2.17\pm 0.01$.

The precise measurement of the power-law slope $p$ (ultimately enabled by the very simple spectral shape) allows us to test with unprecedented accuracy the predictions of the Fermi process for particle acceleration in relativistic shocks.  The power-law  index in trans-relativistic shocks will lie in between the value $p=2$ expected at non-relativistic shock speeds \citep{bell_78,blandford_ostriker_78,blandford_eichler_87} and $p\simeq 2.22$ at ultra-relativistic velocities \citep{kirk_00,achterberg_01,keshet_waxman_05,sironi_13}. From \citet{keshet_waxman_05}, we estimate that the measured $p= 2.17\pm 0.01$ implies a shock Lorentz factor of $\Gamma\sim5$ at 110 d (the $ 3\,\sigma$ c.l. is $\Gamma\sim3-10$). The straightforward implication is then that we are seeing electron acceleration in trans-relativistic shocks in action.\footnote{We remark, though, that a power-law electron spectrum with slope $p$ might not necessarily result in the canonical radiation spectrum $F_\nu \propto \nu^{-(p-1)/2}$, 
if one of the following conditions are met: (\textit{i}) the radiative signature has an appreciable contribution from electrons that cool in the precursor, i.e., upstream of the shock front, which has the effect of hardening the observed spectrum \citep{sironi_spitkovsky_09b,zakine_17}; or (\textit{ii}) the magnetic field self-generated by the shock is not uniform in the post-shock region, but decays away from the shock \citep[e.g.,][]{spitkovsky_08,chang_08,keshet_09,martins_09,haugbolle_10,sironi_13}. In this case, the observed synchrotron spectrum encodes important information on the decay profile of the turbulent post-shock fields \citep{rossi_rees_03,lemoine_12,lemoine_13b}.}

As the non-thermal spectrum of GW170817 showed negligible evolution (Fig. \ref{Fig:SED}), a similar line of reasoning applies to the previous epochs at $t\le 15$ d, from which we conclude that the \emph{observed} non-thermal radiation from GW170817 at $t<115$ d is always dominated by emission from material with relatively small $\Gamma\sim 3-10$. 

These findings are consistent with the picture favored by \cite{Mooley17} (see also \citealt{Salafia17,Kasliwal17,Hallinan17,Nakar18}) of emission from a quasi-isotropic mildly relativistic fireball with stratified ejecta and no surviving ultra-relativistic jet (i.e. their ``choked jet cocoon" scenario), but do not represent a unique prediction from this model as we detail below  (see also \citealt{Nakar18} for an independent study that reached a similar conclusion). A value $\Gamma\sim 3-10$ is significantly smaller than the initial $\Gamma\sim$ a few $100$ inferred for the luminous SGRBs, which are powered by ultra-relativistic jets seen on axis (which have consistently larger inferred values of $p$ \citealt{Fong15}). However, one expects that even a blast wave with large energy $E_{k,iso}\sim 10^{52}\,\rm{erg}$ propagating in a low density medium with $n\sim 10^{-4}-10^{-5}\,\rm{cm^{-3}}$ will have decelerated to $\Gamma\sim 4-5$ by $\sim110$ d since merger, i.e., the shock is mildly relativistic, in excellent agreement with the estimate above based on the physics of particle acceleration at shocks. Current observations are thus also consistent with a scenario where the BNS merger successfully launched an outflow with a collimated ultra-relativistic core (initially pointing away from our line of sight) and less collimated mildly-relativistic wings that dominate the early emission (i.e. the ``successful structured jet" model of Sec. \ref{Sec:StructJet}; \citealt{Jin17,Kathirgamaraju18,Lamb17,LazzatiNature17,Murguia-Berthier17,Troja17GW,Troja18,Davanzo18}). In this latter scenario the emission that we \emph{observe} is also always dominated by radiation from ejecta with relatively small $\Gamma$ at all times. 

We conclude that the observed optically-thin non-thermal spectrum clearly identifies the nature of the emission as synchrotron radiation from a population of electrons accelerated at trans-relativistic shocks with $\Gamma\sim 3-10$. This property, however, is common to both successful structured-jet scenarios and choked-jet scenarios and does \emph{not} identify the nature of the relativistic ejecta. 

\subsection{Off-Axis Relativistic Top-Hat Jets}
\label{Sec:UniformJet}
\begin{figure}
\center
\includegraphics[scale=0.45]{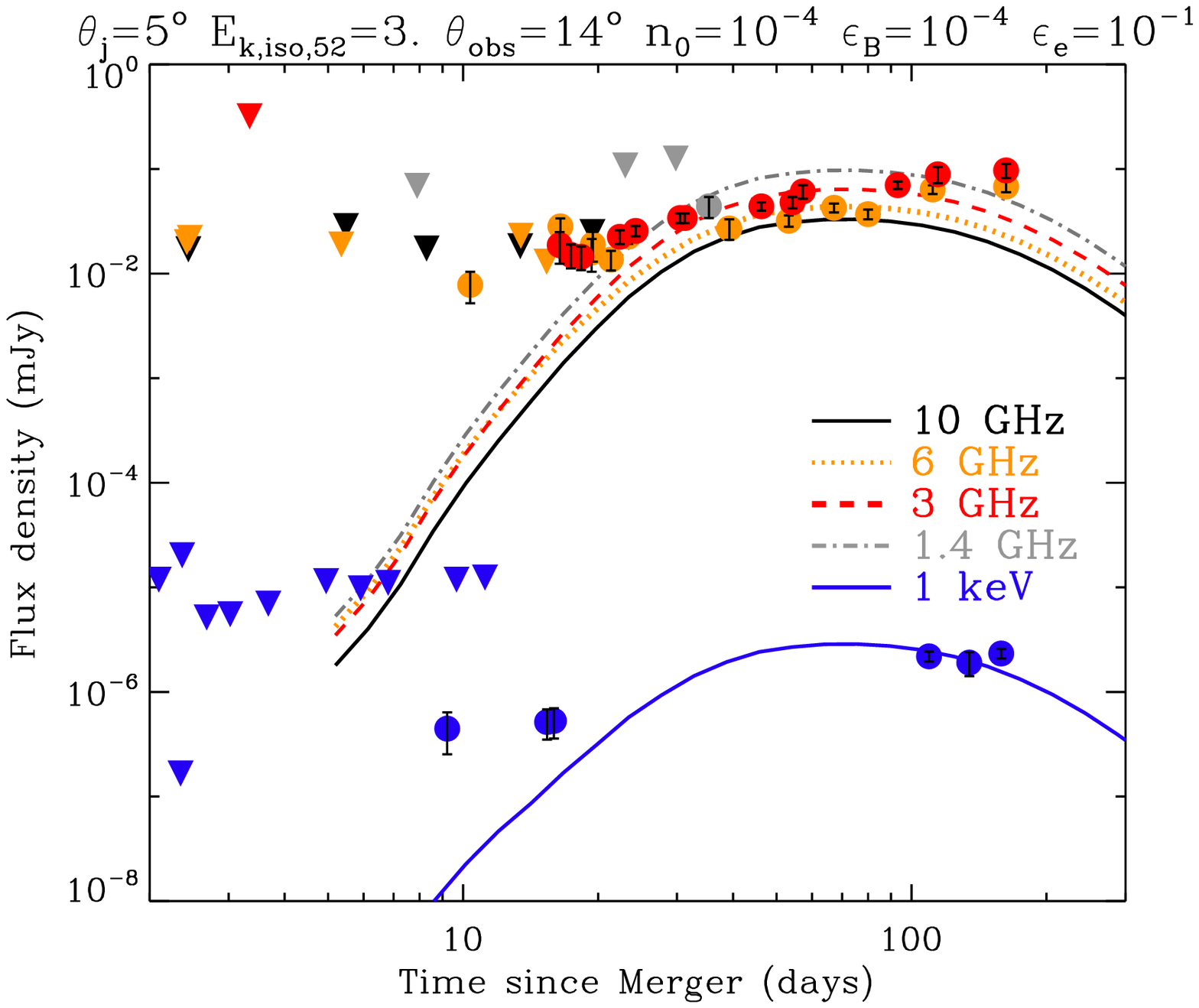}\\
\includegraphics[scale=0.45]{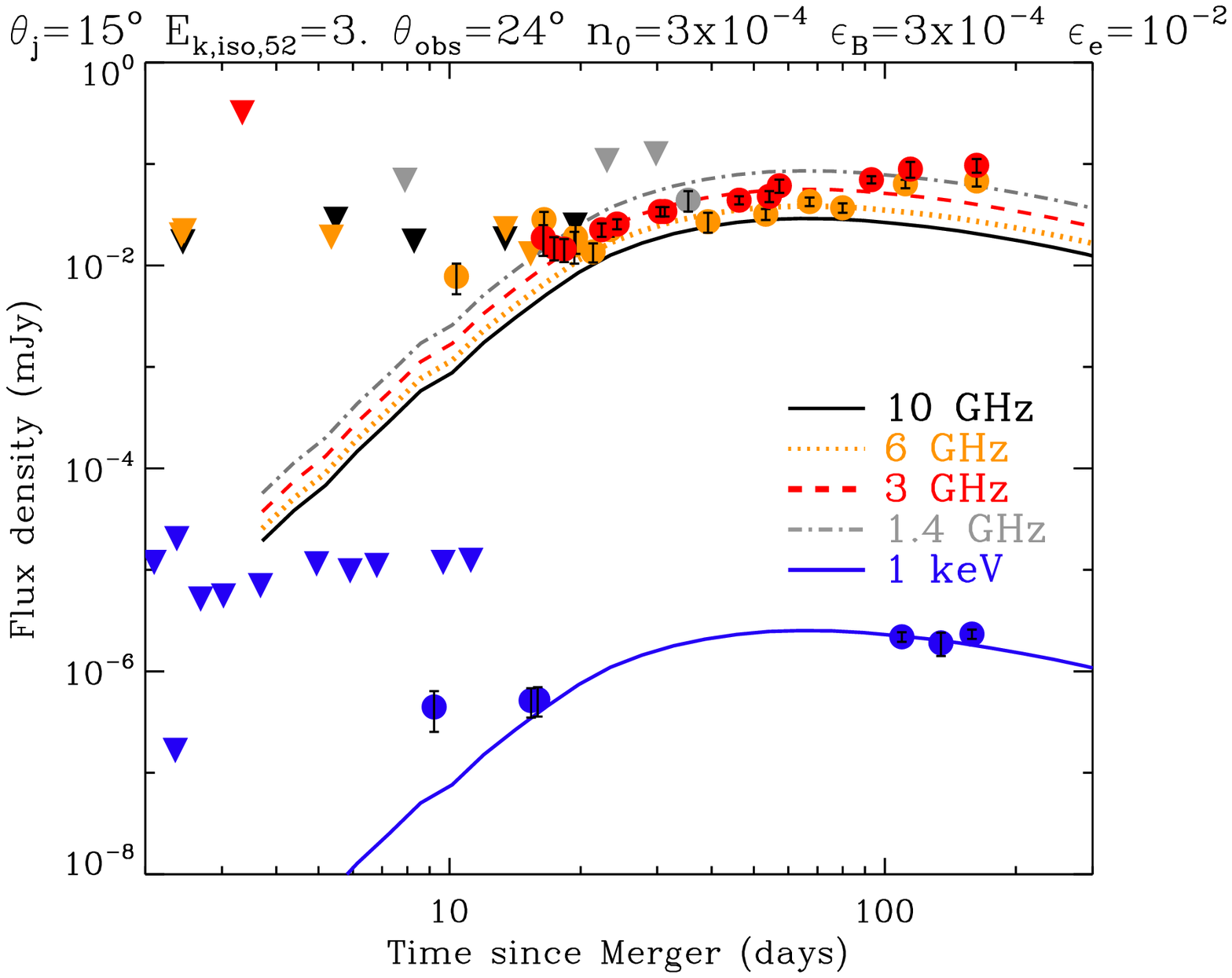}
\caption{Best-fitting top-hat off-axis jet models with $\theta_j=5\degree$ (upper panel) and $\theta_j=15\degree$ (lower panel) for $p=2.1$. These models fail to reproduce observations at early times and do not naturally account for the still-rising light-curve, which is a potential signature of structure $\Gamma(\theta)$ in the jet, with an ultra-relativistic core still out of our line of sight. This is explored and quantified in Sec. \ref{Sec:StructJet}.}
\label{Fig:besttophat}
\end{figure}

\begin{figure*}
\center
\hspace{-1.8 cm}
\includegraphics[scale=0.5]{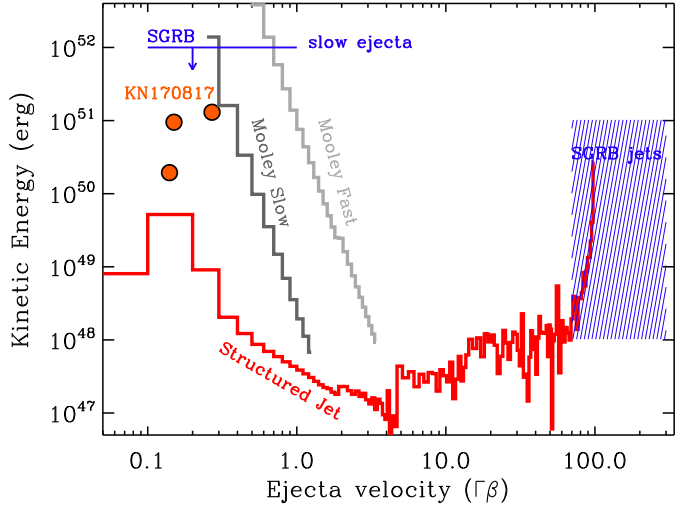}
\caption{Kinetic energy structure of the ejecta of GW170817 for quasi-spherical outflows from \cite{Mooley17} (grey lines) and for  the structured jet that we present here (red line).  Orange filled dots:  kinetic energy of the red, purple and blue kilonova component associated to GW170817 as derived by \cite{Villar17}. Blue lines: SGRBs. For the SGRB slow ejecta we report a representative limit derived from the analysis of very late-time radio observations from \cite{Fong+16}, while the shaded area mark the beaming-corrected $E_k$ of the jet component in SGRBs as derived by \cite{Fong15} for $\epsilon_B=0.1$ (note that smaller values of $\epsilon_B$ would lead to $E_k$  that would extend to larger values, see e.g.  \citealt{Fong15}, their Fig. 7). This plot highlights the difference between quasi-spherical outflows (which lack an ultra-relativistic component and require a large amount of energy to be coupled to slowly moving ejecta $\Gamma<2$) and structured ultra-relativistic outflows (which have properties consistent with SGRBs and can be energetically less demanding).  The peak time of the non-thermal light-curve of GW170817 will constrain the minimum $\Gamma \beta$ of the ejecta in quasi-spherical models. }
\label{Fig:Ek}
\end{figure*}

The late onset of the X-ray and radio emission of GW170817 rules out relativistic jets  with properties similar to those of SGRBs seen on-axis (\citealt{Alexander17GW,Haggard17GW,Hallinan17,Kasliwal17,Margutti17GW,Troja17GW,Mooley17,Ruan17,Granot17off,Fraija17}). Relativistic jets originally pointing away from our line of sight can instead produce rising X-ray and radio emission as they decelerate into the ambient medium (see e.g. \citealt{Granot02b}). 

We first consider top-hat relativistic jets, i.e. jets characterized by a uniform angular distribution of the Lorentz factor within the jet $\Gamma(\theta)$.
This is the simplest jet model and  likely an over simplification of real jets in BNS mergers (e.g. \citealt{Aloy05,Duffell15,Lazzati17pre,Gottlieb18,Kathirgamaraju18}). The simple top-hat jet model is expected to capture the overall behavior of the observed synchrotron emission from relativistic electrons at the shock fronts only after the core of the jet enters into our line of sight, leading to a peak of emission. Before peak, top-hat jets will underpredict the observed emission when compared to structured jets with similar core (Sec. \ref{Sec:StructJet}), i.e. jets with with non-zero $\Gamma(\theta)$ in higher-latitude ejecta at $\theta>\theta_{j}$. 

Figure \ref{Fig:besttophat} shows an update of our modeling of GW170817 with top-hat jets following the same procedure as in \cite{Alexander17GW,Margutti17GW,Guidorzi17} with BOXFIT \citep{vanEerten12b}. We show two representative models for two jet opening angles. Within the top-hat scenario, the most successful models share a preference for low densities $n\sim10^{-4}\,\rm{cm^{3}}$ and large energies $E_{\rm{k,iso}}\sim10^{52}\,\rm{erg}$, with off-axis angles $\theta_{obs}\sim15\degree-25\degree$. As these plots demonstrate,  top-hat jets viewed off-axis fail to reproduce the larger X-ray and radio luminosities of GW170817 at early times $t\lesssim 25$ days and do not naturally account for the mild but steady rise of the non-thermal emission from GW170817. This is expected if the jet in GW170817 has similar core properties as the uniform jets that we are considering here but with  $\Gamma(\theta>\theta_j)>0$ (i.e. a structured jet) and the core of the jet has yet to enter into our line of sight (Sec. \ref{Sec:StructJet}). The X-rays suggest that GW170817 is reaching its peak of emission, which, in this scenario, would imply that the emission from the core of the jet is now close  to entering our line of sight.

In summary, the  failure of the simple top-hat jets motivates the exploration of more realistic structured jets models in Sec. \ref{Sec:StructJet} and should not be interpreted as evidence to discard the notion that GW170817 harbored a fully relativistic outflow directed away from our line of sight.

\subsection{Successful Off-Axis Relativistic Structured Jets}
\label{Sec:StructJet}

Deviation from the simple top-hat jet picture is naturally expected as the relativistic jet has to propagate through the BNS merger immediate environment (e.g. \citealt{Aloy05,Murguia14,Duffell15,Murguia17,Lazzati17pre,Lazzati17post,Kathirgamaraju18,Gottlieb18}), polluted with $\sim0.01\,\rm{M_{\odot}}$ of neutron-rich material that was ejected during the merger (the same material produces the radioactively powered KN, e.g. \citealt{Metzgerhandbook}). Here we consider the scenario where the fully relativistic collimated outflow successfully survived the interaction with the BNS merger ejecta and we refer to this model as successful off-axis relativistic structured jet. In this model the outflow 
has $\Gamma\equiv \Gamma(\theta)$ and $E_{\rm{k,iso}}\equiv E_{\rm{k,iso}} (\theta)$.

This scenario is clearly different from choked-jets, pure-cocoon models and spherical models (favored by  \citealt{Gottlieb17,Hallinan17,Kasliwal17,Mooley17,Salafia17,Nakar18}) where no collimated ultra-relativistic outflow (even when there) survived the interaction with the BNS ejecta. This is clear from Fig. \ref{Fig:Ek}, where we show the $E_{k}$ structure of the two types of outflows. The two classes of models have important implications for the nature of GW170817.  As the emission from the slower jet wings is subdominant at all times when seen on-axis, GW170817 would be consistent with being a canonical SGRB seen from the side, if indeed powered by a successful off-axis structured relativistic jet. GW170817 would be instead a subluminous event and intrinsically different from the population of known SGRBs in the choked-jets and pure-cocoon models. From Fig. \ref{Fig:Ek} it is also clear that quasi-spherical outflows require significantly larger amounts of energy coupled to slow material with $\Gamma\sim1$ ($\gtrsim10^{51}$ erg for the ``fast model" from \citealt{Mooley17}). The quasi-spherical outflows in these models are powered by energy deposited by failed jets. However, observed successful jets  in SGRBs have $\le3\times 10^{50}$ erg (shaded region in Fig. \ref{Fig:Ek}). The two notions can be reconciled only if the most energetic jets never manage to break out, which we find contrived. 

Structured off-axis jets have been specifically discussed in the context of GW170817 by \citealt{Guidorzi17,Kathirgamaraju18,Lamb17,LazzatiNature17,Murguia-Berthier17,Troja17GW,Lyman18,Davanzo18,Troja18,Gottlieb17,Hallinan17,Kasliwal17,Mooley17,Nakar18}. These jets typically have large $E_{\rm{k,iso}}(\theta)$ and $\Gamma(\theta)$ close to the axis of the jet, that decrease for larger angles, resulting in a jet with a narrow, ultra-relativistic core and a wider, mildly relativistic sheath. For off-axis observers, the afterglow is initially dominated by the less collimated emission from the mildly relativistic wings\footnote{This component of emission is missing in top-hat jets, which, as a consequence, show a characteristic $\propto t^2$ rise and underpredict the early time observations as shown in Fig. \ref{Fig:besttophat}.} (which would be also responsible for the detected $\gamma$-ray emission). As time progresses, the jet decelerates, beaming effects become less pronounced and the observer will gradually see the more-luminous, initially ultra-relativistic jet core. 

\begin{figure*}
\center
\includegraphics[scale=0.55]{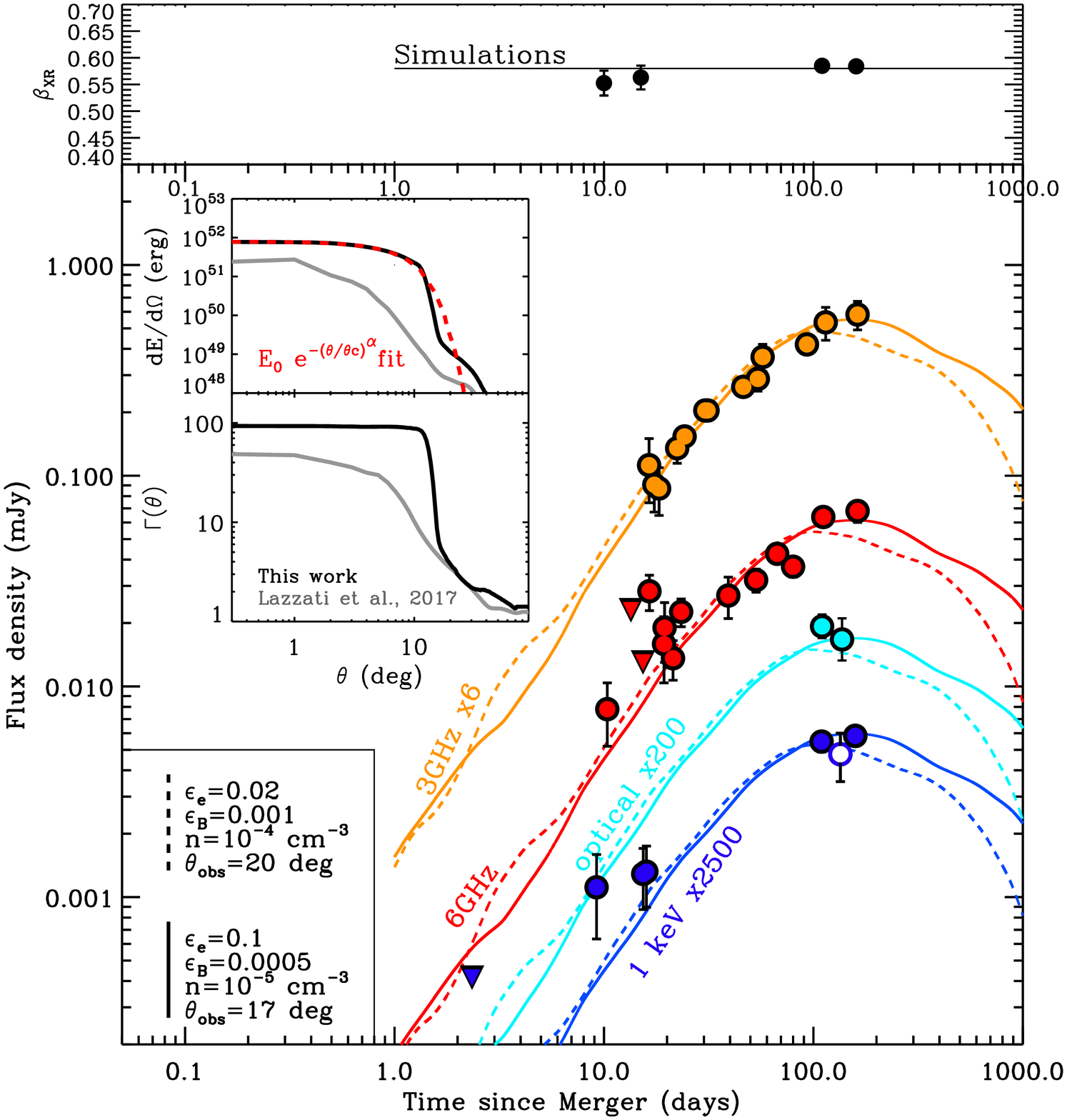}\\
\caption{Results from our simulation of a successful off-axis relativistic jet with structure $\Gamma(\theta)$ and $E_{\rm{iso}}(\theta)$ displayed in the insets, propagating into a low-density environment with $n\sim10^{-5}-10^{-4}\,\rm{cm^{-3}}$ and viewed $\sim20\degree$ off-axis. We use $p=2.16$ and the microphysical parameters reported in the figure. These two representative models can adequately reproduce the current set of observations and predict an optically thin synchrotron spectrum at all times, in agreement with our observations (upper panel). The open blue circle is the XMM X-ray measurement from \cite{Davanzo18}. \emph{Insets}: $E_{\rm{iso}}(\theta)$ and average $\Gamma(\theta)$ from our simulations (black solid lines) at $t=100$ s, compared to the jet structure from \cite{LazzatiNature17} (grey lines). The jet in our simulation has quasi-gaussian structure, with $E_{\rm{iso}}\propto e^{-(\theta/\theta_c)^\alpha}$ and $\alpha\sim 1.9$, $\theta_c\sim9\degree$ (red dashed line). Future observations will be able to constrain the jet-environment parameters.}
\label{Fig:structjet}
\end{figure*}

\begin{figure*}
\center
\includegraphics[scale=0.6]{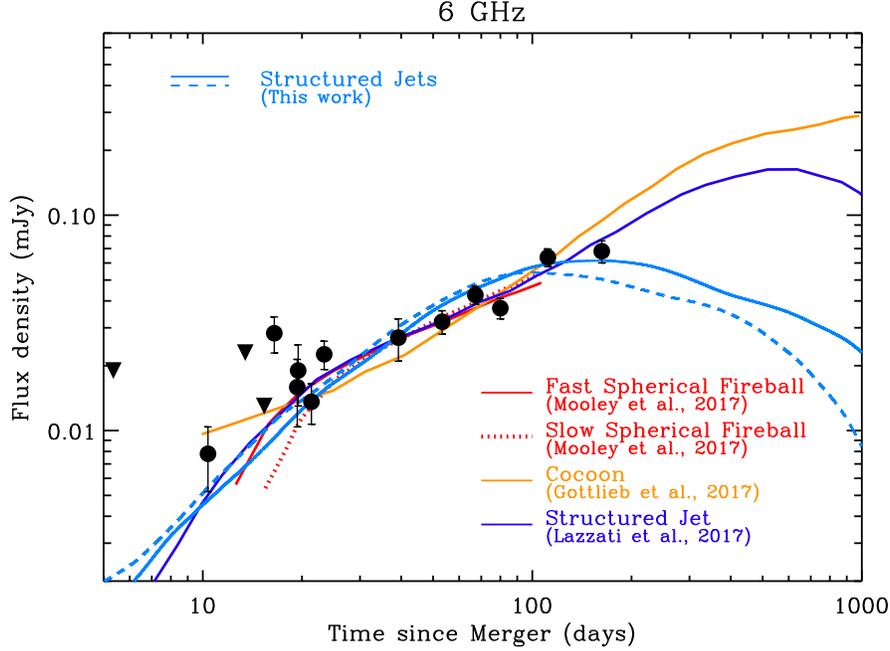}\\
\caption{Comparison of models that fit current observations of GW170817 at radio frequencies (6 GHz). Red and orange lines: quasi-spherical stratified ejecta models from \cite{Mooley17} and cocoon model from \cite{Gottlieb17} where no ultra-relativistic jetted component survived the interaction with the BNS ejecta (i.e. no observer in the Universe observed a regular SGRB associated with GW170817). Blue lines: structured jet models from \cite{LazzatiNature17} (dark blue-line, their best-fitting model) and this work (light-blue lines) where an off-axis ultra-relativistic collimated component is present and contributes to the emission at some point (i.e. GW170817 is consistent with being an ordinary SGRB viewed off-axis). The parameters of our models are the same as in Fig. \ref{Fig:structjet}.  At $t\le 100$ days all the models displayed predict an extremely similar flux evolution (and spectrum), with no hope for current data to distinguish between the two scenarios. The model by \cite{Gottlieb17} and the structured jet model by \cite{LazzatiNature17} predict a continued rise of the radio emission until very late times, and are disfavored by the latest observations at $\sim160$ d, which suggest instead a flattening of the radio light-curve. All off-axis jet models have a similar $\theta_{obs}\sim20\degree$ and the different late-time evolution is a consequence of the different jet-environment parameters.}
\label{Fig:radiomaster}
\end{figure*}

\begin{figure*}
\center
\includegraphics[scale=0.6]{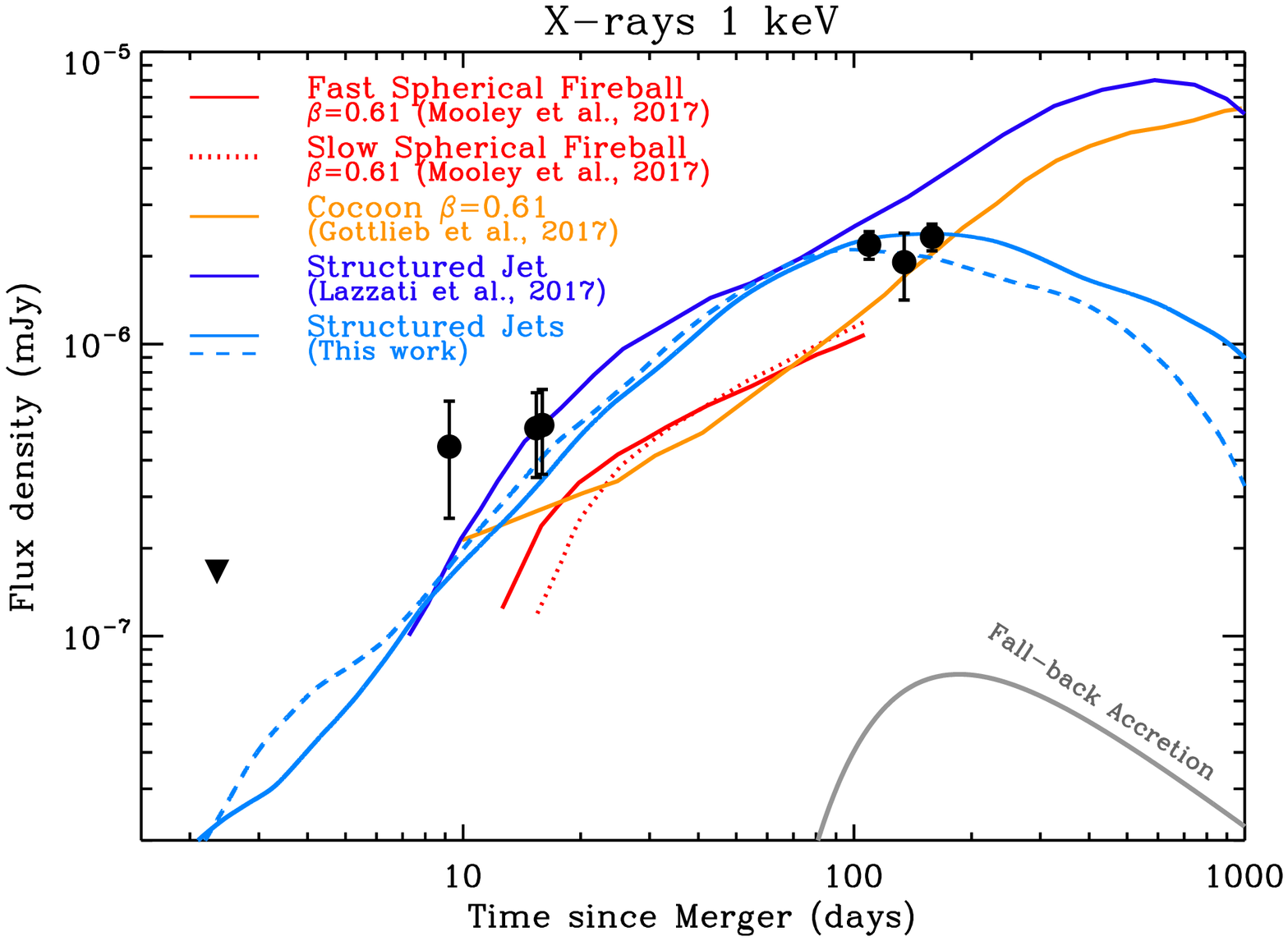}\\ 
\caption{Comparison of successful models at 1 keV. Same color coding as Fig. \ref{Fig:radiomaster}. For the spherical models by  \cite{Mooley17} and \cite{Gottlieb17} we adopt the best fitting spectral index $\beta=0.61$ from \cite{Mooley17} to convert their best fitting radio models into X-rays. These models underpredict the observed X-ray flux. This is a clear indication of a flatter spectral index as we find in Sec. \ref{SubSec:VLA}.  Using $\beta_{XR}\sim0.58$ would bring the models to consistency with the observations. The model by \cite{Gottlieb17} and the structured jet model by \cite{LazzatiNature17} predict a continued rise of the X-ray emission until very late times, and are disfavored by the latest observations at $\sim150$ d, which suggest instead a flattening of the X-ray light-curve. Thick gray line: expected flux from fall-back accretion onto the remnant black hole $F_{\rm{fb}}^{obs}=F_{\rm{fb}}e^{-\tau_X}$ for the fiducial parameters of Sec. \ref{Sec:fallback}.}
\label{Fig:Xraymaster}
\end{figure*}

We use the moving-mesh relativistic hydrodynamics JET code \citep{Duffell13} to simulate the dynamics of explosive outflows launched in neutron star ejecta clouds using an engine model \citep{Duffell15} and density structure similar to \cite{Kasliwal17,Gottlieb17}. We then compute synchrotron light curves from the simulation data using standard synchrotron radiation models (\citealt{Sari98}). We show in Fig. \ref{Fig:structjet} the results for two representative sets of 
jet-environment parameters that successfully account for current observations across the spectrum (a full description of the jet simulations will be presented in Xie et al., in prep.). Specifically, the jet has a narrow ultra-relativistic core of  $\theta_c\sim9\degree$ with $\Gamma\sim100$ surrounded by a mildly relativistic sheath with $\Gamma\sim10$ at $10\degree\lesssim\theta\lesssim60\degree$ (see inset of Fig.  \ref{Fig:structjet}) and propagates in a low-density environments with $n=10^{-5}-10^{-4}\,\rm{cm^{-3}}$. At $t\sim100$ s, the energy in the ultra-relativistic core is $\sim4.4\times 10^{50}$ erg while the sheath carries $\sim1.4\times10^{50}\,\rm{erg}$ (see Xie at al for details).  The observer is located at $\theta_{obs}\sim 17-20\degree$ from the jet axis. We adopt $\epsilon_e=0.02$ ($\epsilon_e=0.1$), $\epsilon_B=0.001$ ($\epsilon_B=0.0005$) with $p=2.16$, within the range of our inferred values (Sec. \ref{SubSec:synch}) for the $n=10^{-4}\,\rm{cm^{-3}}$  ($n=10^{-5}\,\rm{cm^{-3}}$) simulation.
 
 Our model predicts an \emph{observed} broad-band optically thin synchrotron spectrum that extends from the radio to the X-ray band on a $F_{\nu}\propto \nu^{-(p-1)/2}$ spectral segment, from the time of our first observations at $t\sim10$ d until now (at the low densities $n\sim10^{-5}-10^{-4}\,\rm{cm^{-3}}$ favored by our modeling $\nu_c$ is not expected to cross the X-ray band at $t<10^4$ d, see Fig. \ref{Fig:structjet}, upper panel). 
 These findings are consistent with the independent results by \cite{LazzatiNature17} and \cite{Lyman18}, and demonstrate that the persistent optically-thin non-thermal spectrum $F_{\nu}\propto \nu^{-0.585}$ that characterizes GW170817 is not a unique prediction of choked-jets and/or pure-cocoon models. Instead it is a natural expectation from fully-relativistic structured outflows with properties similar to those of SGRBs but viewed from the side. Together with the very similar flux temporal evolution (see Fig. \ref{Fig:radiomaster}-\ref{Fig:Xraymaster}), this makes these two classes of models virtually impossible to distinguish based on current observations.

We compare the results from our simulations to those presented by \cite{LazzatiNature17} in Fig. \ref{Fig:radiomaster}-\ref{Fig:Xraymaster}. The major difference is the flux evolution at $t\ge200$ d, with the \cite{LazzatiNature17} models steadily rising until $t\sim600$ d after merger. As the microphysics parameters ($\epsilon_B=0.002$, $\epsilon_e=0.02$, $p=2.13$) and observing angle ($\theta_{obs}=21\degree$) are very similar to the values of one of our simulations, the different behavior can be ascribed to the combination of possibly different assumptions in the code and a narrower ultra-relativistic core, as shown in the inset of Fig. \ref{Fig:structjet} (which effectively places the observer more off-axis) more slowly decelerating into a lower density environment ($n\sim10^{-5}\,\rm{cm^{-3}}$ vs. $n\sim10^{-4}\,\rm{cm^{-3}}$). In general, outflows with a fully-relativistic core with isotropic energy $\sim 10^{52}\,\rm{erg}$, propagating into environments with $n\le10 ^{-5}\,\rm{cm^{-3}}$ and viewed $\sim20\degree$ off-axis will reach a peak at $t_p\ge 600 $ days ($t_p\sim 2.1\,E_{k,iso,52}^{1/3}\,n^{-1/3}\,((\theta_{obs}-\theta_j)/10\degree)^{8/3}$~days, e.g. \citealt{Granot02}). 

\cite{Gottlieb17,Hallinan17,Kasliwal17,Mooley17,Nakar18} disfavor the structured off-axis model based on circumstantial evidence related to the energetics of the relativistic core needed to power GW170817 compared to SGRBs. We emphasize that these authors do not rule out structured off-axis jets in GW170817 but consider this possibility unlikely based on the large $E_{k,iso}\ge 10^{52}\,\rm{erg}$ required.  We show in Fig. \ref{Fig:Ek} the comparison of the kinetic energies in the different components of the outflow of GW170817 from our simulation with the values inferred for SGRBs from \cite{Fong15}. 
We conclude from this plot that the $E_k$ in the ultra-relativistic ejecta of GW170817 is not unprecedented among SGRBs (shaded blue area, see also \citealt{Fong15}, their Fig. 7) and that GW170817 is consistent with having harbored a normal SGRB directed away from our line of sight. The shaded blue area cover the range of $E_k$ for an assumed $\epsilon_B=0.1$. $E_k$ would extend to larger values for smaller $\epsilon_B=0.01$ (e.g. \citealt{Fong15}, their Fig. 7), thus reinforcing our argument.   In our model the ultra-relativistic component dominates the energetics of the outflow.

Some observational tests to distinguish between the successful structured jet scenario that we support here and the choked-jet/stratified ejecta scenarios have been proposed, including VLBI imaging and the acquisition of a larger sample of GW events with electromagnetic counterparts \citep{Hallinan17,LazzatiNature17,Mooley17}. Here we note that if a collimated outflow of fully relativistic material survived the interaction with the BNS ejecta, the observed light-curve will experience two temporal breaks in the future, which are apparent from Fig. \ref{Fig:structjet} (see also Fig. \ref{Fig:Xraymaster}-\ref{Fig:radiomaster}): a peak when radiation from the jet core  enters the line of sight at $t_p$ (the flattening of the X-ray and radio light-curves is suggesting that GW170817 is approaching its peak of emission), and a jet-break when the far edge of the jet comes into view. In the case of collimated outflows a counter-jet signature is also expected when the jet transitions into the non-relativistic phase at $t_{\rm{NR}}\approx 1100\, (E_{\rm{k,iso,53}}/n)^{1/3}$~days. For $E_{\rm{k,iso}}\ge10^{52}\,\rm{erg}$ and $n\le 10^{-4}\,\rm{cm^{-3}}$ which are relevant here, $t_{\rm{NR}}\ge 30$ yrs  and the appearance of the counter-jet will create a bump in the light-curve at a flux level below the sensitivity of current observing facilities.

\subsection{X-rays from the central compact remnant}
\label{Sec:fallback}

Another source of potential X-ray emission is that originating directly from the central compact remnant, as discussed in detail in  \citep{Murase17}. We first consider an accreting black hole.  The $\approx 2.5M_{\odot}$ black hole created following the merger will still be accreting fall-back debris from the merger event (e.g.~\citealt{Rosswog07,Metzger+10b}).  The accretion luminosity at the present epoch $t$ can be estimated as
\begin{equation}
L_{\rm X,fb} = 0.1\dot{M}_{\rm fb}c^{2} \approx 3\times 10^{38}\,{\rm erg\,s^{-1}}\left(\frac{\dot{M}_{\rm fb}(t=1{\rm s})}{10^{-3}M_{\odot}\,s^{-1}}\right)\left(\frac{t}{120\,{\rm d}}\right)^{-5/3},
\end{equation}
where we have assumed that the fall-back accretion rate follows a $\propto t^{-5/3}$ decay with a value at 1 second post merger normalized to $10^{-3}M_{\odot}\,s^{-1}$ (a characteristic value, which is however uncertain by at least an order of magnitude).  The   $L_{\rm{X,fb}}$ estimated above is thus close to the Eddington luminosity $L_{\rm Edd} \approx 3\times 10^{38}$ erg s$^{-1}$ of the black hole remnant.

The X-ray emission from the central engine is only visible if  not absorbed by the
kilonova ejecta along the line of sight.  Given the estimated ejecta mass of $\gtrsim 10^{-2}M_{\odot}$ and mean velocity $v_{\rm ej} \sim 0.1-0.2$ c (e.g. \citealt{Villar17} for an updated modeling), the optical depth through the ejecta of radius $R \sim v_{\rm ej}t$ and density $\rho \sim M_{\rm ej}/(4\pi R^{3}/3)$ is approximately given by
\begin{eqnarray}
\tau_{\rm X} &\simeq& \rho R \kappa_{\rm X} \nonumber \\ &\approx& 1.2\left(\frac{\kappa_{\rm X}}{10^{3}\rm \,cm^{2}g^{-1}}\right)\left(\frac{M_{\rm ej}}{10^{-2}M_{\odot}}\right)\left(\frac{v_{\rm ej}}{0.2\rm \, c}\right)^{-2}\left(\frac{t}{120 \rm d}\right)^{-2}
\end{eqnarray}
where $\kappa_{\rm X} \sim 10^{3}$ cm$^{2}$ g$^{-1}$ is the expected bound-free opacity of neutral or singly-ionized heavy $r$-process nuclei at X-ray energies $\sim$ a few keV (e.g.~\citealt{Metzgerhandbook}). Thus, depending on the precise ejecta column along our line of sight, we could have $\tau_{\rm X} \lesssim 1$ at the present epoch. Even in the case of negligible opacity to X-ray radiation at the present epoch, $L_{\rm{X,fb}}$ is $\ll$ than the observed X-ray luminosity  $\sim 5\times 10^{39}\,\rm{erg\,s^{-1}}$. The constant radio to X-ray flux ratio over 110 d provides an independent line of evidence against $L_{\rm X,fb}$ dominating the X-ray energy release at late times. Figure \ref{Fig:Xraymaster} shows that $L_{\rm X,fb}$ never dominates the X-ray emission from GW170817. 

\begin{figure}
\center
\includegraphics[scale=0.43]{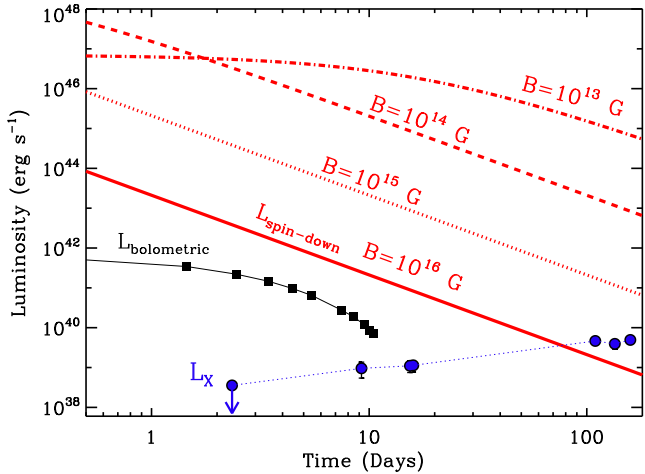}\\
\caption{Red lines: spin-down luminosity for a supramassive NS remnant with magnetic field $B=10^{13}-10^{16}$ G. Black squares: GW170817 bolometric luminosity from \cite{Cowperthwaite17}. Blue filled circles: X-ray luminosity. The spin-down luminosity is always larger than the bolometric energy release from GW170817 at early times, which argues against a long-lived magnetar remnant.}
\label{Fig:spindown}
\end{figure}

We now consider the  spin-down luminosity from  a magnetar remnant as potential source of X-ray radiation at late times. A long-lived magnetar remnant is already disfavored by the KN emission (e.g. \citealt{Cowperthwaite17,Drout17,Kasliwal17,Nicholl17,Pian17,Smartt17,Tanvir17,Villar17}), particularly the inferred presence of lanthanide-rich material created from very neutron-rich ejecta (neutrinos from a long-lived neutron star remnant would transform outflowing neutrons back into protons; see \citealt{MetzgerFernandez14}).  Here we provide an independent argument against the long-lived magnetar scenario. Fig. \ref{Fig:spindown} shows the spin-down luminosity $L_{\rm{sd}}$ for a supramassive NS remnant (Eq. 32-33 from \citealt{Metzgerhandbook}).
At $\sim 10$ d $L_{\rm{sd}}$ greatly exceeds the detected X-ray luminosity for any reasonable magnetic field strength $B\le 10^{17}\,\rm{G}$. However, this argument alone cannot be used to rule out magnetar remnants because at this time $\tau_{\rm X}\gg1$, thus significantly suppressing the X-ray luminosity that  can escape the system and reach the observer, as we showed in \cite{Margutti17GW} (see also Eq. 2 above). \cite{Pooley17} reached the opposite conclusion, as they did not take into account the effects of bound-free opacity from the KN ejecta into their calculations (which, however, is significant). However, as we show in Fig. \ref{Fig:spindown}, the same magnetar engines would produce luminous optical emission at early times \citep{MetzgerPiro14} in excess to the observed bolometric luminosity from GW170817 and for this reason are ruled out.  Finally, one can rule out the formation of a long-lived magnetar in GW170817 by the large rotational energy $\gtrsim 10^{52}$ erg it would have injected into its environment, either into the GRB jet or the kilonova ejecta. As a comparison, in classical SGRBs, long-lived magnetars with rotational energy in the range $\gtrsim 10^{51}-10^{54}$  erg are also ruled out \citep{Fong+16,MargalitMetzger2017}.

We conclude that a central engine origin of the detected X-ray emission is disfavored at all times.

\section{Summary and Conclusions}
\label{Sec:Conc}
Deep Chandra, HST and VLA observations of the BNS event GW170817 $\sim100$ d after merger show a steadily rising emission with $F\propto t^{0.7}\nu^{-0.585}$ across the electromagnetic spectrum, before flattening at $\sim160$ d without showing any sign of spectral evolution. These findings rule out simple models of top-hat jets viewed off-axis (which predict $F\propto t^2$ before peak) and uniform spherical outflows (which predict $F\propto t^3$).  
We use the very simple power-law spectrum extending from the X-rays to the radio band to estimate that the emission is powered by mildly relativistic material with $\Gamma\sim 3-10$. This estimate is solely based on the theory of particle acceleration at shocks (and does not depend on other details of GW170817). 

Models of GW170817 where no ultra-relativistic collimated component survives and the outflow is powered by mildly relativistic stratified ejecta (like those favored by \citealt{Mooley17}) successfully reproduce these observations.\footnote{We note that to reproduce the flattening of the emission within these models it is necessary to introduce a cut into the velocity distribution of the ejecta at some minimum $\Gamma\beta$ value.} Here we offer an alternative interpretation. We employ simulations of the explosive outflows launched in NS ejecta clouds to show that a powerful relativistic core of material can survive the interaction with the BNS ejecta, producing a successful relativistic structured jet (Sec. \ref{Sec:StructJet}). In this case, the \emph{observed} emission is also effectively powered by mildly relativistic ejecta if the ultra-relativistic core is directed away from our line of sight. In this paper we showed one particular model (part of a family of successful models) that fits current observations. A detailed description of the jet simulations using the moving mesh relativistic hydrodynamics code JET \citep{Duffell13} and light curves will be presented in Xie at al, in prep.

A key distinction between the two sets of models is that in the former scenario GW170817 would be intrinsically different from classical SGRBs and the first of a new class of transients. In the latter scenario GW170817 can be instead reconciled with an ordinary SGRB viewed from the side (in SGRBs we are  not sensitive to the presence of lateral structure in the jet as the emission is always dominated by the brighter relativistic core).
Distinguishing between these models is of paramount importance, as it has direct implications on the intrinsic nature of GW170817 and the potential existence of a new class of quasi-spherical transients powered by NS mergers. However, we show here that at the present time the two sets of models predict very similar flux temporal evolution and spectrum.  Observations at $t\ge300$ days, able to track the evolution of $\nu_c$ (which evolves much faster $\propto t^{-2}$ in spherical models, e.g. \citealt{Mooley17}) and to constrain the presence of temporal breaks in the flux evolution are the most promising to discriminate between the two scenarios.

We conclude that current observations do \emph{not} distinguish the nature of the relativistic ejecta and cannot be used to rule out the presence of an off-axis originally ultra-relativistic core of collimated ejecta in the outflow of GW170817. The existence of a new class of BNS merger transients is not required by current observations and GW170817 is consistent with being a classical SGRB viewed off-axis.


\begin{figure*}
\center
\includegraphics[scale=0.5]{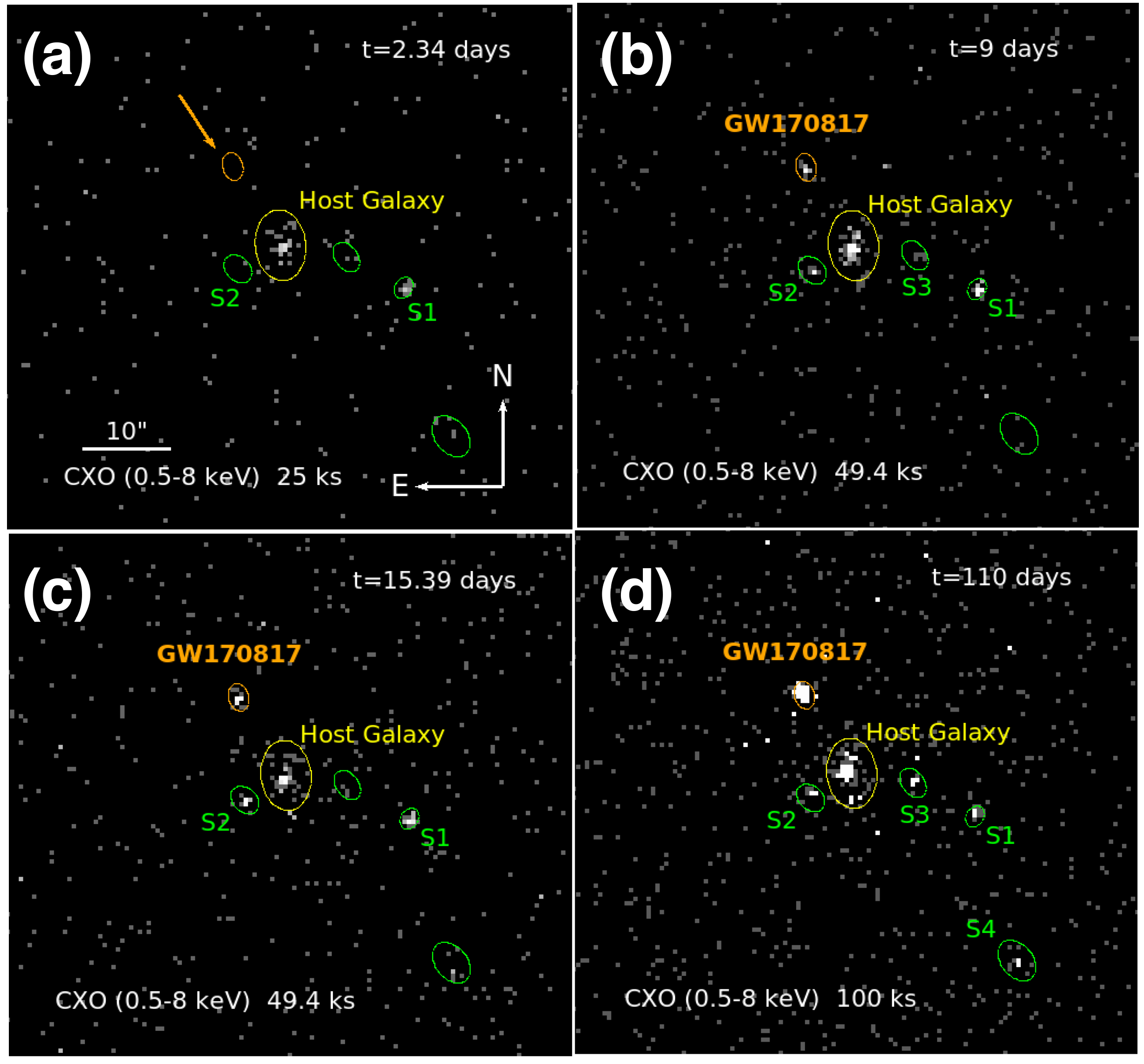}\\
\caption{Evolution of the X-ray emission from GW170817 as seen by the CXO.}
\label{Fig:Xray}
\end{figure*}

\bigskip
RM thanks the entire Chandra team for their work, time and dedication that made these observations possible. RM thanks  P. D'avanzo,  K. Murase, A. Murguia Berthier, E. Nakar, H. Tananbaum for their comments and suggestions to the first version of this work. Support for this work was provided by the National Aeronautics and Space Administration through Chandra Award Number DD7-18096A  issued by the Chandra X-ray Observatory Center, which is operated by the Smithsonian Astrophysical Observatory for and on behalf of the National Aeronautics Space Administration under contract NAS8-03060. WF acknowledges support for Program number HST-HF2-51390.001-A, provided by NASA through a grant from the Space Telescope Science Institute, which is operated by the Association of Universities for Research in Astronomy, Incorporated, under NASA contract NAS5-26555. CG acknowledges University of Ferrara for use of the local HPC facility co-funded by the ``Large-Scale Facilities 2010'' project (grant 7746/2011). AM acknowledges support through NSF grant AST-1715356. We thank University of Ferrara and INFN--Ferrara for the access to the COKA GPU cluster. Development of the Boxfit code was supported in part by NASA through grant NNX10AF62G issued through the Astrophysics Theory Program and by the NSF through grant AST-1009863. The Berger Time-Domain Group at Harvard is supported in part by the NSF through grants AST-1411763 and AST-1714498, and by NASA through grants NNX15AE50G and NNX16AC22G. Simulations for BOXFITv2 have been carried out in part on the computing facilities of the Computational Center for Particle and Astrophysics of the research cooperation ``Excellence Cluster Universe'' in Garching, Germany.  This research was supported in part through the computational resources and staff contributions provided for the Quest high performance computing facility at Northwestern University which is jointly supported by the Office of the Provost, the Office for Research, and Northwestern University Information Technology. We gratefully acknowledge Piero Rosati for granting us usage of proprietary HPC facility. The National Radio Astronomy Observatory is a facility of the National Science Foundation operated under cooperative agreement by Associated Universities, Inc.
\bibliographystyle{apj}
\bibliography{margutti}

\begin{thebibliography}{}

\bibitem[\protect\citeauthoryear{{Abbott} et~al.}{{Abbott}
  et~al.}{2017}]{Abbott17discovery}
{Abbott}, B.~P., et~al. 2017, Physical Review Letters, 119, 161101

\bibitem[\protect\citeauthoryear{{Achterberg} et~al.}{{Achterberg}
  et~al.}{2001}]{achterberg_01}
{Achterberg}, A., {Gallant}, Y.~A., {Kirk}, J.~G.,  \& {Guthmann}, A.~W. 2001,
  \mnras, 328, 393

\bibitem[\protect\citeauthoryear{{Alexander} et~al.}{{Alexander}
  et~al.}{2017}]{Alexander17GW}
{Alexander}, K.~D., et~al. 2017, \apjl, 848, L21

\bibitem[\protect\citeauthoryear{{Aloy}, {Janka}, \& {M{\"u}ller}}{{Aloy}
  et~al.}{2005}]{Aloy05}
{Aloy}, M.~A., {Janka}, H.-T.,  \& {M{\"u}ller}, E. 2005, \aap, 436, 273

\bibitem[\protect\citeauthoryear{{Bell}}{{Bell}}{1978}]{bell_78}
{Bell}, A.~R. 1978, \mnras, 182, 147

\bibitem[\protect\citeauthoryear{{Berger}}{{Berger}}{2014}]{Berger14}
{Berger}, E. 2014, \araa, 52, 43

\bibitem[\protect\citeauthoryear{{Blandford} \& {Eichler}}{{Blandford} \&
  {Eichler}}{1987}]{blandford_eichler_87}
{Blandford}, R.,  \& {Eichler}, D. 1987, \physrep, 154, 1

\bibitem[\protect\citeauthoryear{{Blandford} \& {Ostriker}}{{Blandford} \&
  {Ostriker}}{1978}]{blandford_ostriker_78}
{Blandford}, R.~D.,  \& {Ostriker}, J.~P. 1978, \apjl, 221, L29

\bibitem[\protect\citeauthoryear{{Chang}, {Spitkovsky}, \& {Arons}}{{Chang}
  et~al.}{2008}]{chang_08}
{Chang}, P., {Spitkovsky}, A.,  \& {Arons}, J. 2008, \apj, 674, 378

\bibitem[\protect\citeauthoryear{{Chornock} et~al.}{{Chornock}
  et~al.}{2017}]{Chornock17}
{Chornock}, R., et~al. 2017, \apjl, 848, L19

\bibitem[\protect\citeauthoryear{{Coulter} et~al.}{{Coulter}
  et~al.}{2017}]{Coulter17}
{Coulter}, D.~A., et~al. 2017, ArXiv e-prints, arXiv:1710.05452

\bibitem[\protect\citeauthoryear{{Cowperthwaite} et~al.}{{Cowperthwaite}
  et~al.}{2017}]{Cowperthwaite17}
{Cowperthwaite}, P.~S., et~al. 2017, \apjl, 848, L17

\bibitem[\protect\citeauthoryear{{D'Avanzo} et~al.}{{D'Avanzo}
  et~al.}{2018}]{Davanzo18}
{D'Avanzo}, P., et~al. 2018, ArXiv e-prints

\bibitem[\protect\citeauthoryear{{Drout} et~al.}{{Drout}
  et~al.}{2017}]{Drout17}
{Drout}, M.~R., et~al. 2017, ArXiv e-prints, arXiv:1710.05443

\bibitem[\protect\citeauthoryear{{Duffell} \& {MacFadyen}}{{Duffell} \&
  {MacFadyen}}{2013}]{Duffell13}
{Duffell}, P.~C.,  \& {MacFadyen}, A.~I. 2013, \apj, 775, 87

\bibitem[\protect\citeauthoryear{{Duffell}, {Quataert}, \&
  {MacFadyen}}{{Duffell} et~al.}{2015}]{Duffell15}
{Duffell}, P.~C., {Quataert}, E.,  \& {MacFadyen}, A.~I. 2015, \apj, 813, 64

\bibitem[\protect\citeauthoryear{{Fong} et~al.}{{Fong} et~al.}{2017}]{Fong17GW}
{Fong}, W., et~al. 2017, \apjl, 848, L23

\bibitem[\protect\citeauthoryear{{Fong} et~al.}{{Fong} et~al.}{2015}]{Fong15}
{Fong}, W., {Berger}, E., {Margutti}, R.,  \& {Zauderer}, B.~A. 2015, \apj,
  815, 102

\bibitem[\protect\citeauthoryear{{Fong} et~al.}{{Fong} et~al.}{2016}]{Fong+16}
{Fong}, W., {Metzger}, B.~D., {Berger}, E.,  \& {{\"O}zel}, F. 2016, \apj, 831,
  141

\bibitem[\protect\citeauthoryear{{Fraija} et~al.}{{Fraija}
  et~al.}{2017}]{Fraija17}
{Fraija}, N., {De Colle}, F., {Veres}, P., {Dichiara}, S., {Barniol Duran}, R.,
   \& {Galvan-Gamez}, A. 2017, ArXiv e-prints, arXiv:1710.08514

\bibitem[\protect\citeauthoryear{{Goldstein} et~al.}{{Goldstein}
  et~al.}{2017}]{Goldstein17}
{Goldstein}, A., et~al. 2017, \apjl, 848, L14

\bibitem[\protect\citeauthoryear{{Gottlieb}, {Nakar}, \& {Piran}}{{Gottlieb}
  et~al.}{2018}]{Gottlieb18}
{Gottlieb}, O., {Nakar}, E.,  \& {Piran}, T. 2018, \mnras, 473, 576

\bibitem[\protect\citeauthoryear{{Gottlieb} et~al.}{{Gottlieb}
  et~al.}{2017}]{Gottlieb17}
{Gottlieb}, O., {Nakar}, E., {Piran}, T.,  \& {Hotokezaka}, K. 2017, ArXiv
  e-prints, arXiv:1710.05896

\bibitem[\protect\citeauthoryear{{Granot} et~al.}{{Granot}
  et~al.}{2017}]{Granot17off}
{Granot}, J., {Gill}, R., {Guetta}, D.,  \& {De Colle}, F. 2017, ArXiv
  e-prints, arXiv:1710.06421

\bibitem[\protect\citeauthoryear{{Granot} et~al.}{{Granot}
  et~al.}{2002}]{Granot02b}
{Granot}, J., {Panaitescu}, A., {Kumar}, P.,  \& {Woosley}, S.~E. 2002, \apjl,
  570, L61

\bibitem[\protect\citeauthoryear{{Granot} \& {Sari}}{{Granot} \&
  {Sari}}{2002}]{Granot02}
{Granot}, J.,  \& {Sari}, R. 2002, \apj, 568, 820

\bibitem[\protect\citeauthoryear{{Guidorzi} et~al.}{{Guidorzi}
  et~al.}{2017}]{Guidorzi17}
{Guidorzi}, C., et~al. 2017, \apjl, 851, L36

\bibitem[\protect\citeauthoryear{{Haggard} et~al.}{{Haggard}
  et~al.}{2018}]{Haggard18}
{Haggard}, D., {Nynka}, M., {Ruan}, J.~J., {Evans}, P.,  \& {Kalogera}, V.
  2018, The Astronomer's Telegram, 11242

\bibitem[\protect\citeauthoryear{{Haggard} et~al.}{{Haggard}
  et~al.}{2017}]{Haggard17GW}
{Haggard}, D., {Nynka}, M., {Ruan}, J.~J., {Kalogera}, V., {Cenko}, S.~B.,
  {Evans}, P.,  \& {Kennea}, J.~A. 2017, \apjl, 848, L25

\bibitem[\protect\citeauthoryear{{Haggard et al.}}{{Haggard et
  al.}}{2017}]{HaggardGCNlast}
{Haggard et al.} 2017, GRB Coordinates Network, 22206

\bibitem[\protect\citeauthoryear{{Hallinan} et~al.}{{Hallinan}
  et~al.}{2017}]{Hallinan17}
{Hallinan}, G., et~al. 2017, ArXiv e-prints, arXiv:1710.05435

\bibitem[\protect\citeauthoryear{{Haugb{\o}lle}}{{Haugb{\o}lle}}{2011}]{haugbolle_10}
{Haugb{\o}lle}, T. 2011, \apjl, 739, L42

\bibitem[\protect\citeauthoryear{{Jin} et~al.}{{Jin} et~al.}{2017}]{Jin17}
{Jin}, Z.-P., et~al. 2017, ArXiv e-prints, arXiv:1708.07008

\bibitem[\protect\citeauthoryear{{Kalberla} et~al.}{{Kalberla}
  et~al.}{2005}]{Kalberla05}
{Kalberla}, P.~M.~W., {Burton}, W.~B., {Hartmann}, D., {Arnal}, E.~M.,
  {Bajaja}, E., {Morras}, R.,  \& {P{\"o}ppel}, W.~G.~L. 2005, \aap, 440, 775

\bibitem[\protect\citeauthoryear{{Kasliwal} et~al.}{{Kasliwal}
  et~al.}{2017}]{Kasliwal17}
{Kasliwal}, M.~M., et~al. 2017, ArXiv e-prints, arXiv:1710.05436

\bibitem[\protect\citeauthoryear{{Kathirgamaraju}, {Barniol Duran}, \&
  {Giannios}}{{Kathirgamaraju} et~al.}{2018}]{Kathirgamaraju18}
{Kathirgamaraju}, A., {Barniol Duran}, R.,  \& {Giannios}, D. 2018, \mnras,
  473, L121

\bibitem[\protect\citeauthoryear{{Keshet} et~al.}{{Keshet}
  et~al.}{2009}]{keshet_09}
{Keshet}, U., {Katz}, B., {Spitkovsky}, A.,  \& {Waxman}, E. 2009, \apjl, 693,
  L127

\bibitem[\protect\citeauthoryear{{Keshet} \& {Waxman}}{{Keshet} \&
  {Waxman}}{2005}]{keshet_waxman_05}
{Keshet}, U.,  \& {Waxman}, E. 2005, Physical Review Letters, 94, 111102

\bibitem[\protect\citeauthoryear{{Kim} et~al.}{{Kim} et~al.}{2017}]{Kim17}
{Kim}, S., et~al. 2017, \apjl, 850, L21

\bibitem[\protect\citeauthoryear{{Kirk} et~al.}{{Kirk} et~al.}{2000}]{kirk_00}
{Kirk}, J.~G., {Guthmann}, A.~W., {Gallant}, Y.~A.,  \& {Achterberg}, A. 2000,
  \apj, 542, 235

\bibitem[\protect\citeauthoryear{{Lamb} \& {Kobayashi}}{{Lamb} \&
  {Kobayashi}}{2017}]{Lamb17}
{Lamb}, G.~P.,  \& {Kobayashi}, S. 2017, \mnras, 472, 4953

\bibitem[\protect\citeauthoryear{{Lazzati} et~al.}{{Lazzati}
  et~al.}{2017a}]{Lazzati17post}
{Lazzati}, D., {Deich}, A., {Morsony}, B.~J.,  \& {Workman}, J.~C. 2017a,
  \mnras, 471, 1652

\bibitem[\protect\citeauthoryear{{Lazzati} et~al.}{{Lazzati}
  et~al.}{2017b}]{Lazzati17pre}
{Lazzati}, D., {L{\'o}pez-C{\'a}mara}, D., {Cantiello}, M., {Morsony}, B.~J.,
  {Perna}, R.,  \& {Workman}, J.~C. 2017b, \apjl, 848, L6

\bibitem[\protect\citeauthoryear{{Lazzati} et~al.}{{Lazzati}
  et~al.}{2017c}]{LazzatiNature17}
{Lazzati}, D., {Perna}, R., {Morsony}, B.~J., {L{\'o}pez-C{\'a}mara}, D.,
  {Cantiello}, M.,  \& {Workman}, J.~C. 2017c, ArXiv e-prints, arXiv:1712.03237

\bibitem[\protect\citeauthoryear{{Lemoine}}{{Lemoine}}{2013}]{lemoine_12}
{Lemoine}, M. 2013, \mnras, 428, 845

\bibitem[\protect\citeauthoryear{{Lemoine}, {Li}, \& {Wang}}{{Lemoine}
  et~al.}{2013}]{lemoine_13b}
{Lemoine}, M., {Li}, Z.,  \& {Wang}, X.-Y. 2013, \mnras, 435, 3009

\bibitem[\protect\citeauthoryear{{Lyman} et~al.}{{Lyman}
  et~al.}{2018}]{Lyman18}
{Lyman}, J.~D., et~al. 2018, ArXiv e-prints

\bibitem[\protect\citeauthoryear{{Margalit} \& {Metzger}}{{Margalit} \&
  {Metzger}}{2017}]{MargalitMetzger2017}
{Margalit}, B.,  \& {Metzger}, B.~D. 2017, \apjl, 850, L19

\bibitem[\protect\citeauthoryear{{Margutti} et~al.}{{Margutti}
  et~al.}{2017a}]{Margutti17GW}
{Margutti}, R., et~al. 2017a, \apjl, 848, L20

\bibitem[\protect\citeauthoryear{{Margutti} et~al.}{{Margutti}
  et~al.}{2017b}]{MarguttiGCNlast}
{Margutti}, R., {Fong}, W., {Eftekhari}, T., {Alexander}, K., {Berger}, E.,  \&
  {Chornock}, R. 2017b, The Astronomer's Telegram, 11037

\bibitem[\protect\citeauthoryear{{Margutti et al.}}{{Margutti et
  al.}}{2017}]{MarguttiAtellast}
{Margutti et al.} 2017, Atel, 11037

\bibitem[\protect\citeauthoryear{{Martins} et~al.}{{Martins}
  et~al.}{2009}]{martins_09}
{Martins}, S.~F., {Fonseca}, R.~A., {Silva}, L.~O.,  \& {Mori}, W.~B. 2009,
  \apjl, 695, L189

\bibitem[\protect\citeauthoryear{{McMullin} et~al.}{{McMullin}
  et~al.}{2007}]{McMullin07}
{McMullin}, J.~P., {Waters}, B., {Schiebel}, D., {Young}, W.,  \& {Golap}, K.
  2007, in Astronomical Society of the Pacific Conference Series, Vol. 376,
  Astronomical Data Analysis Software and Systems XVI, ed. R.~A. {Shaw},
  F.~{Hill}, \& D.~J. {Bell}, 127

\bibitem[\protect\citeauthoryear{{Metzger}}{{Metzger}}{2017}]{Metzgerhandbook}
{Metzger}, B.~D. 2017, Living Reviews in Relativity, 20, 3

\bibitem[\protect\citeauthoryear{{Metzger} et~al.}{{Metzger}
  et~al.}{2010}]{Metzger+10b}
{Metzger}, B.~D., {Arcones}, A., {Quataert}, E.,  \& {Mart{\'{\i}}nez-Pinedo},
  G. 2010, \mnras, 402, 2771

\bibitem[\protect\citeauthoryear{{Metzger} \& {Fern{\'a}ndez}}{{Metzger} \&
  {Fern{\'a}ndez}}{2014}]{MetzgerFernandez14}
{Metzger}, B.~D.,  \& {Fern{\'a}ndez}, R. 2014, \mnras, 441, 3444

\bibitem[\protect\citeauthoryear{{Metzger} \& {Piro}}{{Metzger} \&
  {Piro}}{2014}]{MetzgerPiro14}
{Metzger}, B.~D.,  \& {Piro}, A.~L. 2014, \mnras, 439, 3916

\bibitem[\protect\citeauthoryear{{Mooley} et~al.}{{Mooley}
  et~al.}{2017}]{Mooley17}
{Mooley}, K.~P., et~al. 2017, ArXiv e-prints, arXiv:1711.11573

\bibitem[\protect\citeauthoryear{{Murase} et~al.}{{Murase}
  et~al.}{2017}]{Murase17}
{Murase}, K., et~al. 2017, ArXiv e-prints

\bibitem[\protect\citeauthoryear{{Murguia-Berthier} et~al.}{{Murguia-Berthier}
  et~al.}{2014}]{Murguia14}
{Murguia-Berthier}, A., {Montes}, G., {Ramirez-Ruiz}, E., {De Colle}, F.,  \&
  {Lee}, W.~H. 2014, \apjl, 788, L8

\bibitem[\protect\citeauthoryear{{Murguia-Berthier} et~al.}{{Murguia-Berthier}
  et~al.}{2017a}]{Murguia-Berthier17}
{Murguia-Berthier}, A., et~al. 2017a, \apjl, 848, L34

\bibitem[\protect\citeauthoryear{{Murguia-Berthier} et~al.}{{Murguia-Berthier}
  et~al.}{2017b}]{Murguia17}
{Murguia-Berthier}, A., et~al. 2017b, \apjl, 835, L34

\bibitem[\protect\citeauthoryear{{Nakar} \& {Piran}}{{Nakar} \&
  {Piran}}{2018}]{Nakar18}
{Nakar}, E.,  \& {Piran}, T. 2018, ArXiv e-prints

\bibitem[\protect\citeauthoryear{{Nicholl} et~al.}{{Nicholl}
  et~al.}{2017}]{Nicholl17}
{Nicholl}, M., et~al. 2017, \apjl, 848, L18

\bibitem[\protect\citeauthoryear{{Peng} et~al.}{{Peng} et~al.}{2010}]{Peng10}
{Peng}, C.~Y., {Ho}, L.~C., {Impey}, C.~D.,  \& {Rix}, H.-W. 2010, \aj, 139,
  2097

\bibitem[\protect\citeauthoryear{{Pian} et~al.}{{Pian} et~al.}{2017}]{Pian17}
{Pian}, E., et~al. 2017, \nat, 551, 67

\bibitem[\protect\citeauthoryear{{Pooley}, {Kumar}, \& {Wheeler}}{{Pooley}
  et~al.}{2017}]{Pooley17}
{Pooley}, D., {Kumar}, P.,  \& {Wheeler}, J.~C. 2017, ArXiv e-prints,
  arXiv:1712.03240

\bibitem[\protect\citeauthoryear{{Rossi} \& {Rees}}{{Rossi} \&
  {Rees}}{2003}]{rossi_rees_03}
{Rossi}, E.,  \& {Rees}, M.~J. 2003, \mnras, 339, 881

\bibitem[\protect\citeauthoryear{{Rosswog}}{{Rosswog}}{2007}]{Rosswog07}
{Rosswog}, S. 2007, \mnras, 376, L48

\bibitem[\protect\citeauthoryear{{Ruan} et~al.}{{Ruan} et~al.}{2017}]{Ruan17}
{Ruan}, J.~J., {Nynka}, M., {Haggard}, D., {Kalogera}, V.,  \& {Evans}, P.
  2017, ArXiv e-prints, arXiv:1712.02809

\bibitem[\protect\citeauthoryear{{Salafia} et~al.}{{Salafia}
  et~al.}{2017}]{Salafia17}
{Salafia}, O.~S., {Ghisellini}, G., {Ghirlanda}, G.,  \& {Colpi}, M. 2017,
  ArXiv e-prints, arXiv:1711.03112

\bibitem[\protect\citeauthoryear{{Sari}, {Piran}, \& {Narayan}}{{Sari}
  et~al.}{1998}]{Sari98}
{Sari}, R., {Piran}, T.,  \& {Narayan}, R. 1998, \apjl, 497, L17

\bibitem[\protect\citeauthoryear{{Savchenko} et~al.}{{Savchenko}
  et~al.}{2017}]{Savchenko17}
{Savchenko}, V., et~al. 2017, \apjl, 848, L15

\bibitem[\protect\citeauthoryear{{Schlafly} \& {Finkbeiner}}{{Schlafly} \&
  {Finkbeiner}}{2011}]{Schlafly11}
{Schlafly}, E.~F.,  \& {Finkbeiner}, D.~P. 2011, \apj, 737, 103

\bibitem[\protect\citeauthoryear{{Sironi} \& {Spitkovsky}}{{Sironi} \&
  {Spitkovsky}}{2009}]{sironi_spitkovsky_09b}
{Sironi}, L.,  \& {Spitkovsky}, A. 2009, \apjl, 707, L92

\bibitem[\protect\citeauthoryear{{Sironi}, {Spitkovsky}, \& {Arons}}{{Sironi}
  et~al.}{2013}]{sironi_13}
{Sironi}, L., {Spitkovsky}, A.,  \& {Arons}, J. 2013, \apj, 771, 54

\bibitem[\protect\citeauthoryear{{Smartt} et~al.}{{Smartt}
  et~al.}{2017}]{Smartt17}
{Smartt}, S.~J., et~al. 2017, \nat, 551, 75

\bibitem[\protect\citeauthoryear{{Soares-Santos} et~al.}{{Soares-Santos}
  et~al.}{2017}]{Soares-Santos17}
{Soares-Santos}, M., et~al. 2017, \apjl, 848, L16

\bibitem[\protect\citeauthoryear{{Spitkovsky}}{{Spitkovsky}}{2008}]{spitkovsky_08}
{Spitkovsky}, A. 2008, \apjl, 673, L39

\bibitem[\protect\citeauthoryear{{Tanvir} et~al.}{{Tanvir}
  et~al.}{2017}]{Tanvir17}
{Tanvir}, N.~R., et~al. 2017, \apjl, 848, L27

\bibitem[\protect\citeauthoryear{{Troja} \& {Piro}}{{Troja} \&
  {Piro}}{2018}]{Troja18GCN}
{Troja}, E.,  \& {Piro}, L. 2018, The Astronomer's Telegram, 11245

\bibitem[\protect\citeauthoryear{{Troja} et~al.}{{Troja}
  et~al.}{2017a}]{TrojaGCNlast}
{Troja}, E., {Piro}, L., {Ryan}, G., {van Eeerten}, H., {Sakamoto}, T.,  \&
  {Cenko}, S.~B. 2017a, GCN, 22201

\bibitem[\protect\citeauthoryear{{Troja} et~al.}{{Troja}
  et~al.}{2018}]{Troja18}
{Troja}, E., et~al. 2018, ArXiv e-prints

\bibitem[\protect\citeauthoryear{{Troja} et~al.}{{Troja}
  et~al.}{2017b}]{Troja17GW}
{Troja}, E., et~al. 2017b, \nat, 551, 71

\bibitem[\protect\citeauthoryear{{Valenti} et~al.}{{Valenti}
  et~al.}{2017}]{Valenti17}
{Valenti}, S., et~al. 2017, \apjl, 848, L24

\bibitem[\protect\citeauthoryear{{van Eerten}, {van der Horst}, \&
  {MacFadyen}}{{van Eerten} et~al.}{2012}]{vanEerten12b}
{van Eerten}, H., {van der Horst}, A.,  \& {MacFadyen}, A. 2012, \apj, 749, 44

\bibitem[\protect\citeauthoryear{{Villar} et~al.}{{Villar}
  et~al.}{2017}]{Villar17}
{Villar}, V.~A., et~al. 2017, ArXiv e-prints, arXiv:1710.11576

\bibitem[\protect\citeauthoryear{{Williams} et~al.}{{Williams}
  et~al.}{2017}]{pwkit.software}
{Williams}, P.~K.~G., {Clavel}, M., {Newton}, E.,  \& {Ryzhkov}, D. 2017,
  {pwkit: Astronomical utilities in Python}, Astrophysics Source Code Library,
  ascl:1704.001

\bibitem[\protect\citeauthoryear{{Zakine} \& {Lemoine}}{{Zakine} \&
  {Lemoine}}{2017}]{zakine_17}
{Zakine}, R.,  \& {Lemoine}, M. 2017, \aap, 601, A64

\end{thebibliography}

\end{document}